%% file: rho.tex
\documentclass[pre,twocolumn,showpacs,amsmath,amssymb,sort]{revtex4}

\usepackage{graphicx}% Include figure files
\usepackage{dcolumn}% Align table columns on decimal point
\usepackage{bm}% bold math
\usepackage{verbatim}
\usepackage{subfigure}
\usepackage{dcolumn}
\usepackage{xspace}
\usepackage{ifthen}

\newcommand{\pd}{\partial}

\newcommand{\md}{monomer-dimer\xspace}
\newcommand{\dd}{dimer density\xspace}
\newcommand{\p}{\rho}

\newcommand{\combine}{true} 

\begin{document}

%\title{Accurate calculations of an intractable problem -- monomer-dimer model
%in two-dimensional rectangular lattices}
\title{Monomer-dimer model in two-dimensional rectangular lattices
with fixed dimer density}

\author{Yong Kong}
\email{matky@nus.edu.sg}
\affiliation{%
Department of Mathematics\\
National University of Singapore\\
Singapore 117543\\
}%

\date{\today}

\begin{abstract}
The classical monomer-dimer model in two-dimensional lattices
has been shown to belong to the \emph{``\#P-complete''} class,
which indicates the problem is computationally ``intractable''.
We use exact computational method to 
investigate the number of ways to arrange dimers on $m \times n$ 
two-dimensional rectangular lattice strips
with fixed \dd $\rho$. For any \dd $0 < \rho < 1$,
we find a
logarithmic correction term in the finite-size correction of the
free energy per lattice site.
The coefficient of the logarithmic correction term
is exactly $-1/2$. 
This logarithmic correction term is explained by
%This logarithmic correction term fits nicely with 
the newly developed asymptotic theory of Pemantle and Wilson.
%regardless of the parity of the lattice width $n$.
%This is in contrast with the situation reported earlier
%for fixed number of monomers (or vacancies),
%where the logarithmic correction coefficient 
%depends on the number of monomers present and the
%parity of the width of the lattice strip.
The sequence of the free energy 
of lattice strips with cylinder boundary condition
converges so fast that very accurate free energy $f_2(\rho)$ for large lattices
can be obtained. 
For example, for a half-filled lattice,
$f_2(1/2) = 0.633195588930$, while $f_2(1/4) = 0.4413453753046$ 
and $f_2(3/4) = 0.64039026$.
For $\rho < 0.65$, $f_2(\rho)$ is accurate at least to 10 decimal digits.
The function $f_2(\rho)$ reaches the maximum value 
$f_2(\rho^*) = 0.662798972834$ at $\rho^* = 0.6381231$, 
with $11$ correct digits.
This is also the \md constant for two-dimensional rectangular lattices. 
The asymptotic expressions of free energy near close packing 
are investigated for finite and infinite lattice widths.
For lattices with finite width, 
dependence on the parity of the lattice width
is found. For infinite lattices, the data support the functional
form obtained previously through series expansions.
%The data presented are the most accurate results for the \md problem 
%up to date,
%and there are few intractable problems that have been calcualted to
%such high accuracy.

\end{abstract}


%\keywords{monomer dimer}

\maketitle

\section{Introduction}

The \md problem has received much attention not only from  
statistical physics but also from  theoretical computer science.
As one of the classical lattice statistical mechanics models,
the \md model was first used to describe 
the absorption of a binary mixture of molecules of unequal sizes
on crystal surface 
\cite{Fowler1937}.
In the model,
the regular lattice sites are either covered by monomers or dimers.
The diatomic molecules are modeled as rigid dimers 
which occupy two adjacent sites in a regular lattice and no lattice site 
is covered by more than one dimer.  
The lattice sites that are not covered by the dimers
are regarded as occupied by monomers. 
A central problem of the model is to enumerate the 
dimer configurations on the lattice.  
In 1961 an elegant analytical solution was found
for a special case of the problem,
namely when the planar lattice is completely covered by dimers 
(the close-packed dimer problem, or dimer-covering problem) 
\ifthenelse{\equal{\combine}{false}}{ 
\cite{Kasteleyn1961,Temperley1961,Fisher1961}}
{\cite{Kasteleyn1961,Temperley1961}}.
For the general \md problem
where there are vacancies (monomers) in the lattice, 
there is no exact solution.
For three-dimensional lattices,
there is even no exact solution for the special case of 
close-packed dimer problem.
One recent advance is an analytic solution to the 
special case of the problem in two-dimensional lattices
where there is a single 
vacancy at certain specific sites on the boundary 
of the lattice 
\ifthenelse{\equal{\combine}{false}}{ 
\cite{Tzeng2003,Wu2006,Wu2006b}}
{\cite{Tzeng2003,Wu2006}}.
The \md problem also serves as a prototypical problem in the field of
computational complexity 
\ifthenelse{\equal{\combine}{false}}{ 
\cite{Garey1979,Welsh1993,Mertens2002}}  
{\cite{Garey1979}}.  
It has been shown that 
two-dimensional \md problem belongs to 
the \emph{``\#P-complete''} class
and hence is computationally intractable
\ifthenelse{\equal{\combine}{false}}{ 
\cite{Jerrum1987,Jerrum1990}}
{\cite{Jerrum1987}}.

Even though there is a lack of progress in the analytical solution
to the \md problem,
many rigorous results exist, such as
series expansions
\cite{Nagle1966,Gaunt1969,Samuel1980c},
lower and upper bounds on free energy 
\ifthenelse{\equal{\combine}{false}}{ 
\cite{Bondy1966,Hammersley1968,Hammersley1970,Ciucu1998,Lundow2001,FriedlandP05}}
{\cite{Bondy1966,FriedlandP05}},
monomer-monomer correlation function of two monomers
in a lattice otherwise packed with dimers 
\ifthenelse{\equal{\combine}{false}}{ 
\cite{Fisher1963,Hartwig1966}}
{\cite{Fisher1963}},
locations of zeros of partition functions,
\cite{Heilmann1972,Gruber1971},  
and finite-size correction
\ifthenelse{\equal{\combine}{false}}{ 
\cite{Ferdinand1967,IzmailianOH03,Izmailian2005}}
{\cite{Ferdinand1967}}.
Some approximate methods have also been proposed 
\ifthenelse{\equal{\combine}{false}}{ 
\cite{Chang1939,Baxter1968,Nemirovsky1989,LinL94,KenyonRS96,BeichlOS01}}
{\cite{Chang1939,Baxter1968,Nemirovsky1989,LinL94}}.
The \md constant $h_d$ (the exponential growth rate) of the
number of all configurations with different number of dimers
has also been calculated 
\cite{Baxter1968,FriedlandP05}.
By using sequential importance sampling Monte Carlo method,
the dimer covering constant for a three-dimensional cubic lattice
has been estimated \cite{BeichlS99}.
The importance of \md model also comes from the fact that
there is one to one mapping between the Ising model and the \md
model: the Ising model in the absence of an external field is mapped
to the pure dimer model 
\cite{Kasteleyn1963,Fisher1966,Fan1970,McCoy1973},
and the Ising model in the presence of an external field is mapped
to the general \md model \cite{Heilmann1972}.

The major purposes of this paper are (1) to show it is possible
to calculate accurately the free energy of the \md problem
in two-dimensional rectangular lattices at a fixed \dd
by using the proposed computational methods 
(Sections \ref{S:method}, \ref{S:cylinder}, \ref{S:max}, and \ref{S:Baxter}), 
and (2) to use the computational methods
to probe the physical properties of the \md model,
especially at the high \dd limit (Section \ref{S:hdd}).
The high \dd limit is considered to be more difficult
and more interesting than the low \dd limit.
The major result is the asymptotic expression Eq. \ref{E:f_asympt}.
The third purpose of the paper
is to introduce the asymptotic theory of Pemantle and Wilson
\cite{Pemantle2002}, which not only gives a theoretical explanation of
the origin of the logarithmic correction term found by
computational methods reported in this paper (Section \ref{S:lc}), 
but also has the potential to be applicable to other statistical models.

%\subsection{Notation and definitions}
The following notation and definitions will be used throughout the paper.
The configurational grand canonical partition function of the \md system 
in a $m \times n$ two-dimensional lattice is
\begin{equation} \label{E:gpf}
Z_{m,n}(x) = a_{N}(m,n) x^N + a_{N-1}(m,n) x^{N-1} + \cdots + a_{0}(m,n)
\end{equation}
where $a_{s}(m,n)$ is the number of distinct ways to arrange $s$ dimers on the
$m \times n$ lattice, $N = \lfloor mn/2 \rfloor$,
and $x$ can be taken as the activity of a dimer.
The \emph{average}
number of sites covered by dimers (twice the average number of dimers)
of this grand canonical ensemble 
is given by
\begin{equation} \label{E:theta-general}
 \theta_{m,n}(x) = \frac{2}{mn} \frac{ \pd \ln Z_{m,n}(x) } {\pd \ln x}  
 = \frac{2}{mn} 
 \frac{ \sum_{s=1}^N a_s(m,n) s x^s } {\sum_{s=0}^N a_s(m,n) x^s} .
\end{equation}
The limit of this average for large lattices is denoted as $\theta(x)$:
%\begin{equation} \label{E:theta-general-large}
% \theta(x) = \lim_{m,n \rightarrow \infty} \theta_{m,n}(x) .
%\end{equation}  
$\theta(x) = \lim_{m,n \rightarrow \infty} \theta_{m,n}(x)$.
In general we use $\theta_{d} (x)$ 
for the average number of sites covered by dimers
in a $d$-dimensional infinite lattice when the dimer activity is $x$.

The total number of configurations of dimers is given by
$Z_{m,n}(1)$ at $x=1$, and
the \emph{monomer-dimer constant} for a two-dimensional infinite lattice
is defined as
\begin{equation} \label{E:h2_1}
 h_2 = \lim_{m,n \rightarrow \infty} \frac{\ln Z_{m,n}(1)}{mn} . 
\end{equation}
In general, we denote $h_d$ as the monomer-dimer constant
for a $d$-dimensional infinite lattice, 
and $h_d(x)$ as the grand potential per lattice site
at any dimer activity $x$.
For a two-dimensional infinite lattice,
\begin{equation} \label{E:h2-general}
 h_2(x) = \lim_{m,n \rightarrow \infty} \frac{\ln Z_{m,n}(x)}{mn}. 
\end{equation}

In this paper we focus on the number of dimer configurations
at a given \dd $\rho$.
In this sense 
we are working on the \emph{canonical} ensemble.
%where the number of
%dimers $s$ (and hence the number of monomers $mn - 2s$) is fixed.
The connection between the canonical ensemble and
the  grand canonical ensemble
is discussed in Appendix \ref{S:ee}.
We define the \dd for the canonical ensemble as the ratio
\begin{equation} \label{E:rho_0}
 \rho = \frac{2s}{mn}.
\end{equation}
%so that in closed packed lattices, $\rho=1$.
When the lattice is fully covered by dimers, $\rho = 1$.
For a $m \times n$ lattice,
the number of dimers at a given \dd is 
$
 s = \frac{mn \rho}{2}
$.
In the following we use $a_{m,n}(\rho)$ as 
the number of distinct dimer and monomer configurations at the given
\dd $\rho$. 
By using this definition, Eq. \ref{E:gpf}
can be rewritten as
\begin{equation} \label{E:Z_rho}
 Z_{m,n}(x) = \sum_{0 \le \rho \le 1} a_{m,n}(\rho) x^{mn\rho/2}.
\end{equation}

The free energy per lattice site at a given \dd $\rho$ is defined as
\[
 f_{m,n} (\rho) = \frac{\ln a_{m,n}(\rho)}{mn}
\]
and the free energy at a given \dd for a semi-infinite lattice strip
$\infty \times n$ is
\[
 f_{\infty, n}(\rho) = \lim_{m \rightarrow \infty} \frac{\ln a_{m,n}(\rho)}{mn}
 = \lim_{m \rightarrow \infty} f_{m,n} (\rho) .
\]
For infinite lattices where both $m$ and $n$ go to infinity,
%$m, n \rightarrow \infty$:
the free energy is
\[
 f_2(\rho) = f_{\infty, \infty}(\rho) =
 \lim_{m, n \rightarrow \infty} \frac{\ln a_{m,n}(\rho)}{mn}
 = \lim_{n \rightarrow \infty} f_{\infty, n}(\rho).
\]
We use the subscript $d$ in $f_d(\rho)$ to indicate
the dimension of the infinitely large lattice.
From the exact result 
\ifthenelse{\equal{\combine}{false}}{ 
\cite{Kasteleyn1961,Temperley1961,Fisher1961}}
{\cite{Kasteleyn1961,Temperley1961}}
we know $f_2(\rho)$ at $\rho = 1$
\[
 f_2(1) = \frac{G}{\pi} = 0.291560904
\]
where $G$ is the Catalan's constant.
For other values of $\rho > 0$, no analytical result is known,
although several bounds are developed
\ifthenelse{\equal{\combine}{false}}{ 
\cite{Bondy1966,Hammersley1968,Hammersley1970,Ciucu1998,Lundow2001,FriedlandP05}}
{\cite{Bondy1966,FriedlandP05}}.
We will show below that by using the exact calculation method
developed previously
\cite{Kong1999,Kong2006,Kong2006b},
we can calculate $f_2(\rho)$ at an arbitrary \dd $\rho$ with high accuracy. 

The article is organized as follows.
In Section \ref{S:method}, the computational method is outlined.
In Section \ref{S:lc}, we show a logarithmic correction term
in the finite-size correction of $f_{m,n} (\rho)$
for any fixed \dd $0 < \rho < 1$. 
The coefficient of this logarithmic correction term
is exactly $-1/2$, for both cylinder lattices and lattices
with free boundaries.
We give a theoretical explanation for this logarithmic correction term
and its coefficient using the newly developed asymptotic theory
of Pemantle and Wilson \cite{Pemantle2002}.
In this section we point out the universality of this
logarithmic correction term with coefficient of $-1/2$.
This term is not unique to the \md model:
a large class of lattice models has this term
when the ``density'' of the models is fixed.
More discussions of applications of this asymptotic method
to the \md model in particular, and  statistical models
in general, can be found in Section \ref{S:diss}. 
In Section \ref{S:cylinder} we calculate $f_{\infty, n}(\rho)$
on lattice strips $\infty \times n$
for $n=1, \dots, 17$ with cylinder boundary condition.  
The sequence of $f_{\infty, n}(\rho)$
on cylinder lattices converges very fast so that we can obtain
$f_2 (\rho)$ quite accurately.
To the best of our knowledge,
the results presented here are the most accurate
for \md problem in  two-dimensional rectangular lattices
at an arbitrary \dd.
In Section \ref{S:fb} similar calculations of $f_{\infty, n}(\rho)$
are carried out on lattice strips $\infty \times n$ with free boundaries
for $n=1, \dots, 16$. 
Compared with the sequence with cylinder boundary condition,
the sequence $f_{\infty, n}(\rho)$ with free boundaries
converges slower.
In Section \ref{S:max}, the position and values of the
maximum of $f_2(\rho)$ are located: 
$f_2(\rho^*) = 0.662798972834$ at $\rho^* = 0.6381231$.
These results give an estimation of the \md constant 
with $11$ correct digits.
The previous best result is with $9$ correct digits \cite{FriedlandP05}.  
The results are also compared with those obtained
by series expansions and field theoretical methods.
The maximum value of $f_2(\rho)$ is equal to the two-dimensional \md 
constant $h_2$. 
This is one special case of the more general 
relations between the calculated values in the canonical ensemble
and those in the grand canonical ensemble,
and these relations are further discussed
in Appendix \ref{S:ee}. 
In Section \ref{S:Baxter},
the relations developed  in Appendix \ref{S:ee}
are used to compare the results of the computational method
presented in this paper with
those of Baxter \cite{Baxter1968}.
For \md model, the more interesting properties
are at the more difficult high \dd limit.
In Section \ref{S:hdd} asymptotic behavior of the free energy 
$f_{\infty, n}(\rho)$
is examined for high \dd near close packing.
For lattices with finite width, a dependence of the free energy
$f_{\infty, n}(\rho)$
on the parity of the lattice width $n$ is found (Eq. \ref{E:f_m_n}),
consistent with the previous results when the number of monomers is
fixed \cite{Kong2006b}.
The combination of the results in this section and those
of Section \ref{S:lc} leads to the asymptotic expression 
Eq. \ref{E:f_asympt} for near close packing \dd.
The asymptotic expression of $f_2(\rho)$, 
the free energy on an infinite lattice,
is also investigated near close packing.
The results support the functional forms obtained
previously through series expansions \cite{Gaunt1969},
but quantitatively the value of the exponent is lower than previously
conjectured.
In Appendix \ref{S:1d} we put together in one place
various explicit formulas for the one-dimensional lattices ($n=1$).
These formulas can be used to check the formulas developed
 for the more general
situations where $n > 1$.
As an illustration, an explicit application of the 
Pemantle and Wilson asymptotic method is also given for $n=1$.
%
%
%
%In the Appendices, extensive data are listed.
%
%
%

\section{Computational methods} \label{S:method}

The basic computational strategy is to use exact calculations
to obtain a series of partition functions $Z_{m,n}(x)$ 
of lattice strips $m \times n$.
Then for a given \dd $\p$, $f_{m, n}(\p)$ can be calculated
using arbitrary precision arithmetic.
By fitting $f_{m, n}(\p)$ to a given function (Sections \ref{S:cylinder},
\ref{S:fb}, and \ref{S:hdd}),
$f_{\infty, n}(\p)$ can be estimated with high accuracy.
From $f_{\infty, n}(\p)$, $f_2(\p)$ can then be estimated
using the special convergent properties of the sequence
$f_{\infty, n}(\p)$ on the cylinder lattice strips
(Section \ref{S:cylinder}).

\subsection{Calculation of the partition functions} \label{SS:pf}

The computational methods used here have been described in details
previously \cite{Kong1999, Kong2006, Kong2006b}.
The full partition functions (Eq. \ref{E:gpf})
are calculated recursively for lattice strips on cylinder lattices and lattices
with free boundaries.
As before, 
all calculations of the terms $a_{s}(m,n)$ 
in the partition functions use exact integers, 
and when logarithm is taken to calculate free energy $f_{m,n}(\rho)$,
the calculations are done with precisions much higher than the
machine floating-point precision.
The bignum library used is GNU MP library (GMP)
for arbitrary precision arithmetic
(version 4.2) \cite{gmp}.
The details of the calculations on lattices with free boundaries can be found
in Ref. \onlinecite{Kong2006}, so in the following only information
on cylinder lattices is given.

For a $m \times n$ lattice strip,  
a square matrix $M_n$ is set up based on two rows of the lattice strip
with proper boundary conditions.
The vector $\Omega_m$, which consists of partition function of Eq. \ref{E:gpf}
as well as other contracted partition functions \cite{Kong1999},
is calculated by the following recurrence
\begin{equation} \label{E:M}
  \Omega_m = M_n \Omega_{m-1}.
\end{equation}
Similar recursive method
has also been used for other
combinatorial problems, such as calculation of the number of
independent sets \cite{Calkin1998}. 
For a cylinder lattice strip, the matrix $M_n$ is constructed 
in a similar way as that with free boundaries \cite{Kong2006}.
The total valid number ($v_c(n)$) and unique number ($u_c(n)$) of 
configurations are given respectively by the generating function
$\sum_n v_c(n) x^n = x(3+x-x^3)/(1-3x-x^2)/(x-1)/(x+1)$ 
and the formula
\[
 u_c(n) = \sum_{ d \backslash n } \frac{\varphi(d) 2^{n/d}}{2n} + 
 \begin{cases}
   2^{(n-1)/2}         & \text{if $n$ odd} \\
   2^{n/2-1}+2^{n/2-2} & \text{if $n$ even}
 \end{cases}
\] 
where $\varphi(m)$ is Euler's totient function,
which gives the number of integers relatively prime to integer $m$.
The size of matrix $M_n$ is $u_c(n) \times u_c(n)$.
The first $17$ terms of the sequence $v_c(n)$ are: 
$3$, $10$, $36$, $118$, $393$, $1297$, $4287$, $14158$, 
$46764$, $154450$, $510117$, $1684801$, 
$5564523$, $18378370$, $60699636$, $200477278$, and $662131473$.
The first $17$ terms of the sequence $u_c(n)$ are: 
$2$, $3$, $4$, $6$, $8$, $13$, $18$, $30$, $46$, 
$78$, $126$, $224$, $380$, $687$, $1224$, $2250$, and $4112$.
It is noted that the sequence $u_c(n)$ is exactly the same as that
shown in column 2, Table 1 of Ref. \onlinecite{FriedlandP05}.
Calculations based on the dominant eigenvalues of the matrices
of the cylinder lattice strips for $n=4$, $6$, $8$, and $10$
are carried out by Runnels \cite{Runnels1970}.
The sizes of $M_n$ for cylinder lattice strips are smaller
when compared with the corresponding numbers for lattice strips
with free boundaries \cite{Kong2006},
which allows for calculations on wider lattice strips.  
For cylinder lattice strips, full partition functions
are calculated for $n=1, \dots, 17$,
with length up to $m=1000$ for $n=1, \dots, 13$,
$m=880$ for $n=14$,
$m=669$ for $n=15$,
$m=474$ for $n=16$, and
$m=325$ for $n=17$.
The corresponding numbers for lattice strips with free boundaries
are reported in Ref. \onlinecite{Kong2006}.

\subsection{Interpolation for arbitrary \dd $\p$} \label{SS:idd}

In this paper the main quantity to be calculated is $f_2(\rho)$.
The starting point of the calculations is the full partition function
Eq. \ref{E:gpf} for different values of $n$ and $m$.
Finite values of $n$ and $m$
only lead to discrete values of \dd $\p$, as defined in
Eq. \ref{E:rho_0}.  
For example, when $n=7$ and $m=11$, the number of dimers $s$
takes the values of $0, 1, \dots, 38$,
and \dd $\p$ of this lattice can only be one of the following values: 
$0, 2/77, \dots, 76/77$. 
In general, for fixed finite $m$ and $n$,
$\p$ can only be a rational number:
\[
 \rho = \frac{p}{q}
\]
where $p$ and $q$ are positive integers.
When $\rho$ is expressed as a rational number, the number of dimers
is given by 
\begin{equation} \label{E:rho}
 s = \frac{mn \rho}{2} = \frac{mnp}{2q}.
\end{equation}
This expression is only meaningful
if $mnp$ can be divided by $2q$.
When we write the grand canonical partition function in the form
of Eq. \ref{E:Z_rho} for finite $m$ and $n$,
we implicitly imply that Eq. \ref{E:rho} is satisfied.

In the following we use the rational \dd
$\rho = p/q$ whenever possible so that 
the value of $a_{m,n}(\p)$ can be read directly from 
the partition function of $m \times n$ lattices.
Depending on the values of $p$ and $q$,
some dimer densities, such as $\p=1/2$,  can be realized in many lattices,
while others can only be realized in small number of lattices
with special combinations of values of $m$ and $n$.
%In general, the bigger $p$ and $q$,
%the fewer lattices that can realize the $\p$.
In many situations 
it becomes impossible to use rational $\rho$. 
For example,
in Section \ref{S:max} 
the location of the maximum of $f_2(\rho)$
is searched within a very small region of $\rho$,
and in Section \ref{S:Baxter}, in order to compare the results from different
methods,
$\rho$ takes the output values of other
computational methods \cite{Baxter1968}.
In such situations,
if the rational form of $\p$ were used,
$p$ and $q$ would become so big that not enough data points
which satisfy Eq. \ref{E:rho} could be found for the fitting
in the $m \times n$ lattice strip.
To calculate $f_{m,n} (\rho)$ 
for an arbitrary real number $\p$ ($0 < \rho < 1$),
interpolation of the exact data points is needed.
Since full partition functions have been calculated for 
fairly long lattice strips, proper interpolation procedure
can yield highly accurate values of $f_{m,n} (\rho)$ 
for an arbitrary real number $\rho$.
For interpolation, we use the standard 
Bulirsch-Stoer rational function interpolation method 
\cite{Stoer1980,Press1992}.
For any real number $\rho$, Eq. \ref{E:rho} is used to calculate
the corresponding number of dimers $s$, which may not be an integer. 
On each side of this value of $s$, $30$ exact values of
$a_{s} (m,n)$ are used (if possible) in the interpolation.
If on one side there are not enough  exact data points of $a_{s} (m,n)$,
extra data points on the other side of $s$ are used 
to make the total number of
exact data points as $60$. 
For the high \dd case (Section \ref{S:hdd}),
the total number of data points used is changed to $30$.
We also take care that no extrapolation is used: 
if $\p$ is greater than the maximum \dd for a given $m \times n$ lattice,
the data point from this lattice is not used.
Let's look at the above example of $11 \times 7$ lattice again.
%As for the above example of the $11 \times 7$ lattice,
For this lattice,
the highest \dd is $76/77$.
If calculation is done for a \dd $\p=0.99$, 
since $\p=0.99$ is greater than $76/77 \approx 0.987$, 
the data point from this lattice will not be used in the following steps
to avoid inaccuracy introduced by unreliable extrapolations.

\subsection{Fitting procedure} \label{SS:fit}

The fitting experiments are carried out by using the ``fit'' function
of software {\sc gnuplot} (version 4.0) \cite{gnuplot40}
on a 64-bit Linux system. 
%(kernel 2.6.12-1.1381\_FC3smp)
%with dual AMD Opteron processors.
The fit algorithm implemented is the nonlinear least-squares
(NLLS) Levenberg-Marquardt method \cite{Marquart1963}.
All fitting experiments use the default value $1$ as the initial value 
for each parameter, and each fitting experiment is done
independently.
As done previously \cite{Kong2006,Kong2006b},
only those $a_{m,n} (\rho)$ with $m \ge 100$
are used in the fitting.
Since $a_{m,n} (\rho)$ is calculated for relatively long lattice strips
(in the $m$ direction, see Section \ref{SS:pf}), 
the estimates of $f_{\infty, n}(\p)$
are usually quite accurate, up to $12$ or $13$ decimal place.
The accuracy for this fitting step 
is limited by the machine floating-point precision, 
since {\sc gnuplot} uses machine floating-point representations, 
instead of arbitrary precision arithmetic.
We would have used the GMP library to implement a fitting program
with arbitrary precision arithmetic.
%as we did in the previous steps.
This would increase the accuracy in the estimation of $f_2(\p)$
when $\p$ is small.
For the major objective of this paper, i.e., to investigate the behavior
of $f_2(\p)$ when $\p \rightarrow 1$ (Section \ref{S:hdd}), 
however, the current accuracy is adequate.
At high \dd limit,
the convergence of $f_{\infty, n}(\p)$ towards $f_{2}(\p)$ 
is much slower than
at low \dd limit.
With lattice width $n \le 17$ used for the current calculations,
$f_{\infty, n}(\p)$ is far from converging to the machine 
floating-point precision when $\p \rightarrow 1$.

\section{Logarithmic corrections of the free energy
at fixed \dd} \label{S:lc}

For lattice strips $m \times n$ with a fixed width $n$ and
a given \dd $\rho$,
the coefficients $a_{m,n}(\rho)$ of the partition functions 
are extracted to fit the following function:
\begin{equation} \label{E:fit}
%\begin{multline} \label{E:fit}
 f_{m,n} (\rho) =
 \frac{\ln a_{m,n}(\rho)}{m n}
 = c_0 + \frac{c_1}{m} + \frac{c_2}{m^2} + 
 \frac{c_3}{m^3} + \frac{c_4}{m^4} + \frac{\ell}{n} \frac{\ln(m+1)}{m}
\end{equation}
%\end{multline}
where $c_0 = f_{\infty, n} (\rho)$.

For both cylinder lattices and lattices with free boundaries,
the fitting experiments clearly show that
$\ell = -1/2$, accurate up to at least six decimal place,
for any \dd $0 < \rho < 1$. 
This result holds for both odd $n$ and even $n$.
This is in contrast with the results reported earlier 
for the situation with a fixed number of monomers (or vacancies),
where the logarithmic correction coefficient 
depends on the number of monomers present and the
parity of the width of the lattice strip 
\cite{Kong2006,Kong2006b}.
We notice that a coefficient $-1/2$ also appears 
in the logarithmic correction term
of the free energy studied in Ref. \onlinecite{Tzeng2003},  
which is a special case of the \md problem where there is a single 
vacancy at certain specific sites on the boundary 
of the lattice.

For the general \md model,
to our best knowledge,
this logarithmic correction term with coefficient of exactly $-1/2$
has not been reported before in the literature.
The recently developed multivariate asymptotic theory
by Pemantle and Wilson \cite{Pemantle2002},
however, gives an explanation of this term and its coefficient.
This theory applies to combinatorial problems when the 
multivariate generating function of the model is known.
For univariate generating functions,
asymptotic methods are well known and have been used for
a long time. The situation is quite different for 
multivariate generating functions.
Until recently, techniques to get asymptotic expressions from 
multivariate generating functions were ``almost entirely missing''
 (for review, see Ref. \onlinecite{Pemantle2002}).
The newly developed Pemantle and Wilson method applies to a large class of  
multivariate generating functions in a systematic way.
In general the theory applies to generating functions with multiple variables,
and for the bivariate case that we are interested in here,
the generating function of two variables takes the form
\begin{equation} \label{E:F}
 F(x, y) = \frac{G(x, y)}{H(x, y)} = \sum_{s, m = 0}^{\infty} a_{sm} x^s y^m
\end{equation}
where $G(x, y)$ and $H(x, y)$ are analytic, and $H(0,0) \ne 0$.
In this case, Pemantle and Wilson method gives the asymptotic expression
as
\begin{equation} \label{E:asympt}
 a_{sm} \sim \frac{G(x_0, y_0)} {\sqrt{2\pi}} x_0^{-s} y_0^{-m} 
 \sqrt{\frac{-y_0 H_y(x_0, y_0)}{m Q(x_0, y_0)}}
\end{equation}
where $(x_0, y_0)$ is the positive solution to the two equations
%\begin{align} \label{E:asympt_dir}
%  H(x, y)       &= 0 \\
%  s x \frac{ \pd{H} } {\pd {x}}  &= m y \frac{ \pd{H} } {\pd {y}} \notag
%\end{align}
\begin{equation} \label{E:asympt_dir}
  H(x, y)       = 0, \qquad
  m x \frac{ \pd{H} } {\pd {x}}  = s y \frac{ \pd{H} } {\pd {y}} 
\end{equation}
and $Q(x, y)$ is defined as
\begin{widetext}
\[
 -(x H_x) (y H_y)^2 -(y H_y) (x H_x)^2
 - [(y H_y)^2 (x^2 H_{xx}) + (x H_x)^2 (y^2 H_{yy}) 
   - 2 (x H_x) (y H_y)(xy H_{xy})]. 
\]
\end{widetext}
Here $H_x$, $H_y$, etc. are partial derivatives $\pd{H}/\pd{x}$,
$\pd{H}/\pd{y}$, and so on.
One of the advantages of the method over previous ones 
is that the convergence of Eq. \ref{E:asympt} is \emph{uniform} 
when $s/m$ and $m/s$ are bounded.

For the \md model discussed here, 
with $n$ as the finite width  of the lattice strip,
$m$ as the length,
and $s$ as the number of dimers, 
we can construct the bivariate generating function $F(x, y)$ as
\begin{equation} \label{E:bgf}
 F(x, y) = \sum_{m=0}^{\infty} \sum_{s=0}^{mn/2} a_s(m,n) x^s y^m 
 = \sum_{m=0}^{\infty} Z_{m,n}(x) y^m. 
\end{equation}
For the \md model, as well as a large class of lattice models
in statistical physics,
the bivariate generating function $F(x, y)$ 
is always in the form of Eq. (\ref{E:F}),
with $G(x, y)$ and $H(x, y)$ as polynomials in $x$ and $y$.
In fact, we can get $H(x, y)$ directly from matrix
$M_n$ in Eq. (\ref{E:M}).
It is closely related to the characteristic function of $M_n$ \cite{Kong1999}:
$H(x, y) = \det(M_n - I/y) \times y^w  $, where $w$ is the size of the
matrix $M_n$.
As an illustration, the bivariate generating function 
$F(x, y)$ for the one-dimensional lattice ($n=1$)
is shown in Eq. \ref{E:F_1} of Appendix \ref{S:1d}.

When the \dd is fixed, which is the case discussed here,
$s = \rho m n /2$.
If we substitute this relation into Eq. (\ref{E:asympt_dir}),
then we see that the solution $(x_0, y_0)$ of Eq. (\ref{E:asympt_dir}) 
only depends on $\rho$ and $n$, and does not depend on $m$ or $s$. 
Substituting this solution $(x_0(\rho), y_0(\rho))$ into Eq. (\ref{E:asympt})
we obtain
\begin{widetext}
\begin{equation} \label{E:asympt_md}
 f_{m,n} (\rho) \sim -\frac{1}{n} \ln(x_0^{\rho n /2} y_0) 
 -\frac{1}{2}\frac{\ln m}{mn} 
 + \frac{1}{mn} \ln \left( G(x_0, y_0) 
\sqrt{\frac{-y_0 H_y(x_0, y_0)}{2\pi Q(x_0, y_0)}}
 \right).
\end{equation}
\end{widetext}
From this asymptotic expansion we obtain the 
logarithmic correction term with coefficient of $-1/2$ exactly,
for any value of $n$. 
In fact, this asymptotic theory predicts that there exists
such a logarithmic correction term with coefficient of $-1/2$
for a large class of lattice models 
when the two variables involved are proportional, that is, 
when the models are at fixed ``density''. 
For those lattice models which can be described
by bivariate generating functions,
this logarithmic correction term with coefficient of $-1/2$
is universal when those models are at  fixed ``density''.
For the \md model, this proportional relation is for
$s$ and $m$ with $s = \rho m n /2$.
An explicit calculation for $n=1$ is shown in Appendix \ref{S:1d}.

For a fixed \dd $\rho$ and a fixed lattice width $n$, 
the first term of Eq. \ref{E:asympt_md} is
a constant and does not depend on $m$.
we identify it as $f_{\infty,n} (\rho)$
\begin{equation} \label{E:asympt_rho}
 f_{\infty,n} (\rho) = -\frac{1}{n} \ln(x_0^{\rho n /2} y_0) .
\end{equation}

In all the following fitting experiments, we set $\ell = -1/2$ 
for Eq. \ref{E:fit}.

\section{Cylinder lattices} \label{S:cylinder}

For the \md problem at a given \dd $\rho$ in cylinder lattice strips,
the sequence $f_{\infty, n}(\rho)$ converges very fast to 
$f_2(\rho)$, as can be seen from a few sample data in Table \ref{T:c}.
In the table, values of $f_{\infty, n}(\rho)$ for $\rho = 1/4$, $1/2$,
$3/4$, and $1$ are listed.
Two obvious features can be observed: 
(1) The function $f_{\infty, n}(\rho)$ is an increasing function of odd $n$,
but a decreasing function of even $n$. 
Furthermore, 
for finite integer values of $h$ and $k$,
\begin{equation} \label{E:inequ}
f_{\infty, 2h}(\rho) > f_2(\rho) > f_{\infty, 2k+1}(\rho).
\end{equation}
The value $f_{\infty, n}(\rho)$ oscillates around the limit value
$f_2(\rho)$ 
from even $n$ to odd $n$. 
(2) The smaller the value of $\rho$, the faster the rate of convergence
of $f_{\infty, n}(\rho)$ towards $f_2(\rho)$.
Rational values of $\rho$ are used for the calculations in Table \ref{T:c}
and no interpolation of $a_{m,n}(\rho)$ is used.
The numbers of data points used in the fitting are listed in the
parentheses.

As a check of the accuracy of the results, the data at $\rho=1$
can be compared with the exact solution.
For a cylinder lattice strip $\infty \times n$,
the exact expression for $f_{\infty, n}(1)$ reads
as \cite{Kasteleyn1961}
\begin{equation} \label{E:a_0_e}
  f_{\infty, n}(1) = \frac{1}{n} \ln 
%\left[ 
  \prod_{i=1}^{n/2}
  \left[ \sin \frac{(2i - 1)\pi}{n}
  + \left(1 + \sin^2 \frac{(2i - 1)\pi}{n} \right)^{\frac{1}{2}} 
  \right] 
%\right]
.
\end{equation}
The results from Eq. \ref{E:a_0_e} are listed as the last column
in Table \ref{T:c}.
As mentioned in the previous Section,
all input data are exact integers and the logarithm of these integers
is taken with high precision before the fitting. 
The only places where accuracy can be lost are in the fitting procedure
as well as the approximation introduced by the fitting function
 Eq. \ref{E:fit}.
Comparisons of the data in the last two columns
of Table \ref{T:c} show that,
as far as the fitting procedure is concerned,
the calculation accuracy is up to $12$ or $13$ decimal place. 

Another check for the accuracy of the fitting procedure
is through the exact expression Eq. \ref{E:r1} of one-dimensional
strip ($n=1$) at various \dd $\rho$. 
The data are listed in the first row of Table \ref{T:c}.
By using Eq. (\ref{E:asympt_rho}) 
of the Pemantle and Wilson asymptotic method, 
we can also compare
the fitting results with exact asymptotic values for small values
of $n$ (data not shown).
All these checks confirm consistently 
that the accuracy of the fitting procedure
is up to $12$ or $13$ decimal place.
See Section \ref{S:diss} for further discussions on this issue.

The fast convergence of $f_{\infty, n}(\rho)$ and
the property of Eq. \ref{E:inequ}
make it possible to obtain
$f_2(\rho)$ quite accurately, especially when $\rho$ is not too close
to $1$. Some of the values of $f_2(\rho)$ at 
rational $\rho = p/q$ for small $p$ and $q$
are listed in Table \ref{T:list}.
As in Table \ref{T:c}, no interpolation of $a_{m,n}(\rho)$ is used. 
The numbers in square brackets indicate the next digits
for $n=16$ (upper bound) and $n=17$ (lower bound).
The data show that when $\rho \le 0.65$,
the $f_2(\rho)$ is accurate up to at least $10$ decimal place.
It should be pointed out that the data listed are just raw data,
showing digits that have already converged
for $n=16$ and $n=17$.
If the pattern of convergence of these raw data is explored 
and extrapolation technique is used, as is done in Section \ref{S:max},
it is possible to get even more correct digits.
As shown in Section \ref{S:max}, the true value of
$f_2(\rho)$ is not the average of $f_{\infty, 16}(\rho)$
and $f_{\infty, 17}(\rho)$. Instead, it should lie closer
to $f_{\infty, 17}(\rho)$.
%
%
%
%More data are listed in the tables in the Appendices.
%
%
%

%\begin{comment}

\begingroup
\squeezetable
\begin{table*}
\caption{
The coefficient $c_0$ ($f_{\infty, n}(\rho)$)
for different $n$ and $\rho$ on cylinder lattice strips $\infty \times n$.
The numbers in parentheses are the number of data points used in the fitting.
The first row for $n=1$ is from the exact expression Eq. \ref{E:r1}.
The last column is from the exact expression Eq. \ref{E:a_0_e}
when $\rho = 1$.
Rational $\rho$ is used here and no interpolation of $a_{m,n}(\rho)$ is used.
\label{T:c}}
\begin{ruledtabular}
\begin{tiny}
\begin{tabular}{llllll}
\input{latex_table_v2.1}

\end{tabular}
\end{tiny}
\end{ruledtabular}
\end{table*}
\endgroup

%\end{comment}

\begin{table}
\caption{
List of $f_2(\rho)$ for different $\rho$.
Numbers in square brackets indicate the next digits
for $n=16$ (upper bound) and $n=17$ (lower bound).
Rational $\rho$ is used here and no interpolation of $a_{m,n}(\rho)$ is used.
\label{T:list}}
\begin{ruledtabular}
\begin{tabular}{ll}
$\rho$  & $f_2(\rho)$           \\
\hline
0	& 0			\\
1/20	& 0.1334362263587	\\	
1/10	& 0.229899144084[8..9]	\\	
3/20	& 0.310823643168[1..2]	\\	
1/5	& 0.380638530252[1..2]	\\	
1/4	& 0.4413453753046	\\	
3/10	& 0.4940275921700	\\	
1/3	& 0.525010031447[5..6]	\\	
7/20	& 0.539305666744[5..6]	\\
2/5	& 0.5775208675757	\\
9/20	& 0.6088200746799	\\
1/2	& 0.633195588930[4..5]	\\
11/20	& 0.650499726669[5..8]	\\
3/5	& 0.66044120984[2..5]	\\
13/20	& 0.6625636470[2..4]	\\
2/3	& 0.661425713[7..8]	\\
7/10	& 0.65620036[0..1]	\\
3/4	& 0.64039026[3..5]	\\
4/5	& 0.6137181[3..4]	\\
17/20	& 0.573983[2..3]	\\
9/10	& 0.51739[1..2]		\\	
19/20	& 0.435[8..9]		\\
1	& 0.29[0..3]		\\

\end{tabular}
\end{ruledtabular}
\end{table}

\section{Lattices with free boundaries} \label{S:fb}

We also carry out similar calculations for lattice strips
on $m \times n$ two-dimensional lattices with free boundaries,
for $n = 2, \dots, 16$.
A few sample data are shown in Table \ref{T:fb1}.
In the table, values of $f_{\infty, n}^{\text{fb}}(\rho)$ 
for $\rho = 1/4$, $1/2$,
$3/4$, and $1$ are listed.
The data in Table \ref{T:fb1} show that the sequence of 
$f_{\infty, n}^{\text{fb}}(\rho)$ in lattices with free boundaries
converges slower than that in cylinder lattices.
Furthermore, in contrast to the situation in cylinder lattices,
$f_{\infty, n}^{\text{fb}}(\rho)$ is an increasing function of $n$
for $0 < \rho < 1$: $f_{\infty, n}^{\text{fb}}(\rho)$ approaches $f_2(\rho)$
(the same value as that for cylinder lattices) monotonically from below.
When $\rho = 1$, the functions $f_{\infty, 2k}^{\text{fb}}(1)$ 
and $f_{\infty, 2k+1}^{\text{fb}}(1)$
are increasing functions, with 
$f_{\infty, 2k}^{\text{fb}}(1) > f_{\infty, 2k-1}^{\text{fb}}(1)$
and $f_{\infty, 2k}^{\text{fb}}(1) > f_{\infty, 2k+1}^{\text{fb}}(1)$.
Due to the slow convergence rate and the lack of property like 
Eq. \ref{E:inequ},
it is difficult to obtain $f_2(\rho)$ reliably from the data
on the lattice strips with free boundaries.

As we did in the previous Section, we also take advantage of
the known exact solution for $\rho=1$ as a check for the 
numerical accuracy of the fitting procedure.
The exact result for lattice strips with free boundaries is given by
\cite{Kasteleyn1961}
\begin{equation} \label{E:a_0_e_fb}
  f_{\infty, n}^{\text{fb}}(1) = \frac{1}{n} \ln \left[ \prod_{i=1}^{\frac{n}{2}}
  \left( \cos \frac{i \pi}{n+1}
  + \left(1 + \cos^2 \frac{i \pi}{n+1} \right)^{\frac{1}{2}} 
  \right) \right].
\end{equation}
The last column in Table \ref{T:fb1} lists the values given by 
Eq. \ref{E:a_0_e_fb}, which can be compared with the calculated values
from the fitting experiments in the column next to it.
Again, as shown in the previous Section, the 
accuracy at $\rho=1$ is up to $11$ or $12$ decimal place for most of the values
of $n$.

%\begin{comment}

\begingroup
\squeezetable
\begin{table*}
\caption{
The coefficient $c_0$ ($f_{\infty, n}^{\text{fb}}(\rho)$)
for different $n$ and $\rho$ on lattice strips $\infty \times n$
with free boundaries.
The numbers in parentheses are the number of data points used in the fitting.
The last column is from the exact expression Eq. \ref{E:a_0_e_fb}
when $\rho = 1$.
Rational $\rho$ is used here and no interpolation of $a_{m,n}(\rho)$ is used.
\label{T:fb1}}
\begin{ruledtabular}
\begin{tiny}
\begin{tabular}{llllll}

\input{latex_table_fb_v2.1}

\end{tabular}
\end{tiny}
\end{ruledtabular}
\end{table*}
\endgroup

%\end{comment}

\section{Maximum of free energy and the \md constant} \label{S:max}

It is well known that $f_d(\rho)$ is a continuous concave function of $\rho$
and at certain \dd $\rho^*$, $f_d(\rho)$ reaches its maximum
\cite{Hammersley1966}. However, there is no analytical knowledge of the
location ($\rho^*$) and value ($f_d(\rho^*)$) of the maximum for $d > 1$.
As is shown in Appendix \ref{S:ee}, 
the maximum of $f_d(\rho)$
is equal to the \md constant: $f_d(\rho^*) = h_d$.
Currently the best value for $h_2$ is given in Ref. \onlinecite{FriedlandP05},
which gives $h_2 = 0.6627989727 \pm
                   0.0000000001 $,
with $9$ correct digits.
The location of the maximum, $\rho^*$, is controversial.
Baxter gives the value of $\rho^* = 0.63812311$ \cite{Baxter1968},
while Friedland and Peled state, ``it is reasonable to assume
that the value $p^*$, for which $\lambda_2(p^*) = h_2$, is fairly close to
$p(2) = \frac{9-\sqrt{17}}{8}  \approx 0.6096118$'' 
(Here the original notation is used:
$p$ is our $\rho$ and $\lambda_2(p)$ is our $f_2(\rho)$) \cite{FriedlandP05}.

\begin{figure}
  \centering
  \includegraphics[angle=270,width=\columnwidth]{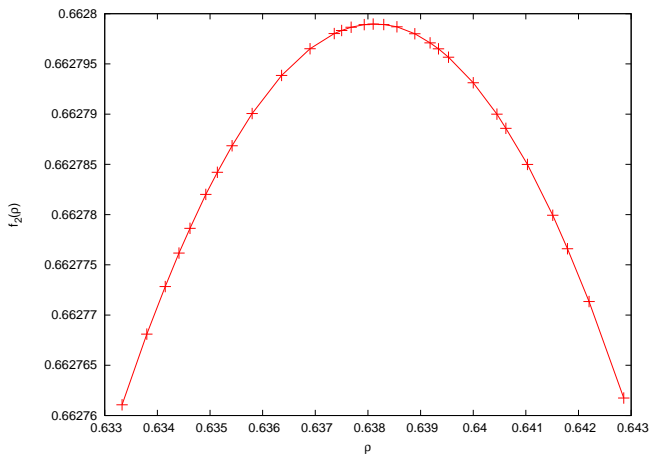}
  \caption{(Color online)
    The function of $f_2(\rho)$ in the region of $19/20 \le \rho \le 9/14$.
    All data points use rational $\rho$, so no interpolation is used.
    \label{F:max_1}}
\end{figure}

In this Section
we use the same computational procedure described in the previous
sections to locate accurately the maximum of $f_2(\rho)$.
Using rational \dd $\rho = p/q$ and choose appropriate $p$ and $q$, 
we can locate the maximum to a fairly small region,
as shown in Figure~\ref{F:max_1}.

\begin{figure}
  \centering
  \includegraphics[angle=270,width=\columnwidth]{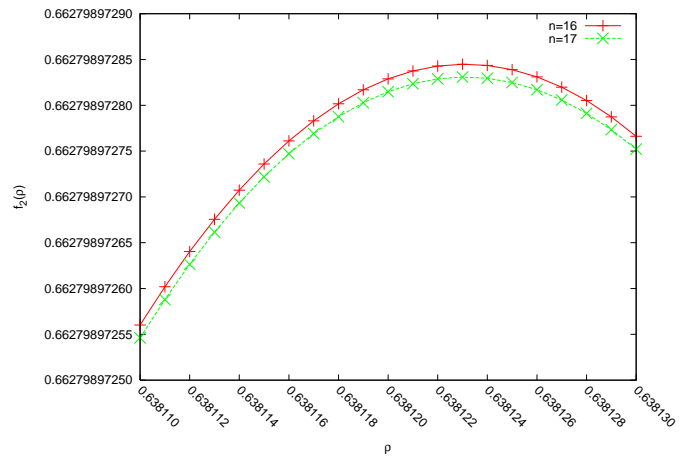}
  \caption{(Color online)
    The function of $f_{\infty, n}(\rho)$ 
    in the region of $0.638110 \le \rho \le 0.638130$ for $n=16$ (upper curve)
    and $n=17$ (lower curve). 
    \label{F:max_2}}
\end{figure}

\begin{figure}
  \centering
  \includegraphics[angle=270,width=\columnwidth]{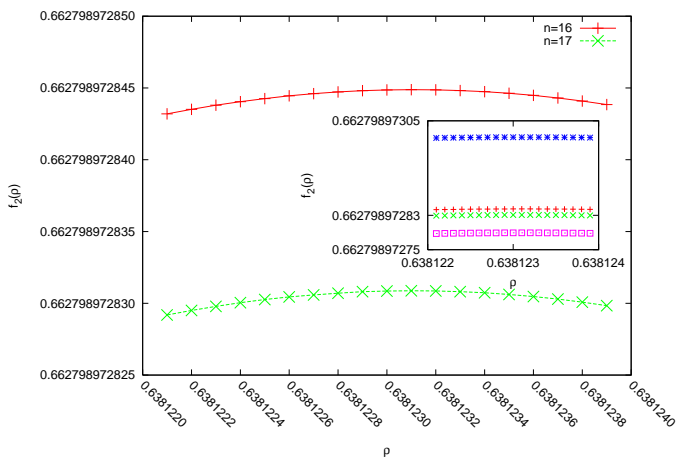}
  \caption{(Color online)
    The function of $f_{\infty, n}(\rho)$ 
    in the region of $0.6381221 \le \rho \le 0.6381239$ for $n=16$ 
    (upper curve)
    and $n=17$ (lower curve).
    In the inset the data points from $n=14$ and $n=15$ are also shown.
    From top to bottom in the inset:
    $n=14$, $n=16$, $n=17$, and $n=15$.
    \label{F:max_3}}
\end{figure}

\begin{figure}
  \centering
  \includegraphics[angle=270,width=\columnwidth]{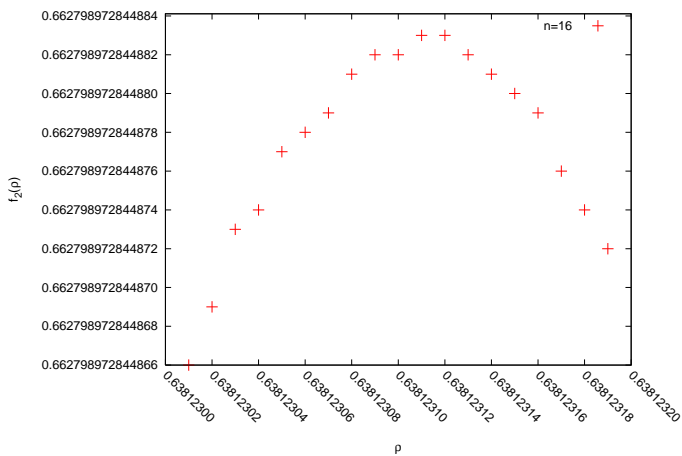}
  \caption{(Color online)
    The function of $f_{\infty, n}(\rho)$ 
    in the region of $0.63812301 \le \rho \le 0.63812319$ for $n=16$. 
    \label{F:max_4}}
\end{figure}

With the interpolated data for $a_{m,n} (\rho)$, 
we can locate $\rho^*$ and $f_2(\rho^*)$ more accurately.
As shown in  Figures \ref{F:max_2} and \ref{F:max_3},
we find that 
\[
0.662798972831 < f_2(\rho^*) < 0.662798972845,
\] 
where the value of $f_{\infty, n}(\rho)$ for $n=16$ is used as the 
upper bounds, and that for $n=17$ as the lower bounds.
From Figures \ref{F:max_2}, \ref{F:max_3} and \ref{F:max_4}
we can locate $\rho^*$ as
\[
0.63812310 < \rho^* <  0.63812312.
\]

The values of $f_{\infty, n}(\rho)$ around $\rho^*$ are listed in 
Table \ref{T:max}.
Inspection of the convergent rate of these data for even and odd values
of $n$ suggests that for both sequences, the convergent rate is geometric.
If we assume that
\begin{equation} \label{E:geometry}
 f_{\infty, n}(\rho) = f_2(\rho) - \alpha \eta^n,
\end{equation}
then the data points
at $\rho = 0.63812311$ of $n=12$, $14$, and $16$ can be used
to obtain an extrapolated value of $f_2(\rho) = 0.6627989728336$,
while the data points of $n=13$, $15$, and $17$ give another
extrapolation value $f_2(\rho) = 0.6627989728341$. 
Together these two extrapolation values converge to 
$f_2(\rho^*) = 0.662798972834$, 
with $11$ correct digits.

We can also get the same conclusion graphically from 
Figure \ref{F:max_3}.
By inspecting the pattern of the data points of different values of $n$
in the inset of Figure \ref{F:max_3},
we notice that the difference between the data points of $n=14$ and $n=16$
is bigger than the difference between $n=15$ and $n=17$.
This indicates that the true value of $f_2(\rho^*)$
lies closer to the data point of $n=17$ than the data point of $n=16$.
From Figure \ref{F:max_3} we are quite sure that the $11$-th
digit of $f_2(\rho^*)$ is $3$ instead of $4$,
and the $12$-th digit is probably $4$, as indicated by the
two extrapolation values mentioned above.

The value of $f_2(\rho^*)$ is in excellent agreement with that 
reported in Ref. \onlinecite{FriedlandP05}, 
which gives $9$ correct digits 
(Friedland and Peled also guess correctly the $10$-th digits as $8$). 
The value  also agrees with that in Ref. \onlinecite{Baxter1968},
which gives $8$ correct digits \cite{FriedlandP05}.
The value of $\rho^*$ is exactly that of Baxter \cite{Baxter1968}.

By using field theoretical method, Samuel uses the following
relation to transform the activity $x$ into a new variable $\omega$
\cite{Samuel1980c}
\begin{equation} \label{E:sam}
 x = \frac{\omega}{(1-4\omega)^2}.
\end{equation}
This relation is very close to the one used by Nagle \cite{Nagle1966}.
By substituting this relation into Gaunt's series expansions \cite{Gaunt1969},
Samuel obtained new series for various lattices, including the
rectangular lattice (Eq. (5.12) of Ref. \onlinecite{Samuel1980c}).
The value of \md constant in two-dimensional rectangular lattice
can be calculated at $x=1$ by using his
series as $0.66279914$. As we can see, this only gives
five correct digits. 
Nagle used the following transform \cite{Nagle1966}
\begin{equation} \label{E:nagle}
 x = \frac{\omega}{(1-3\omega)^2}.
\end{equation}
By using Gaunt's series, a value of $0.6627988$ is obtained, with
six correct digits. 

%
%
%
%More data for Figure \ref{F:max_3} are listed in Appendix \ref{A:max}.
%
%
%
\begingroup
\squeezetable
\begin{table*}
\caption{
The coefficient $c_0$ ($f_{\infty, n}(\rho)$)
for different $n$ and $\rho$ on cylinder lattice strips $\infty \times n$
around $\rho^*$.
The numbers in parentheses are the number of data points used in the fitting.
\label{T:max}}
\begin{ruledtabular}
\begin{tabular}{lllll}

\input{latex_table_v6.1}

\end{tabular}
\end{ruledtabular}
\end{table*}
\endgroup

To conclude this section, we compare our results on the maximum of $f_2(\rho)$
with the approximate
formulas of Chang \cite{Chang1939} and Lin and Lai \cite{LinL94}.
The Chang's approximate formula is given below
%\begin{multline*}
\[
 f_2^C(\rho) = 
 -\frac{1}{2} [4 \ln 4 - (4-\rho) \ln(4-\rho) 
%\\
   + \rho \ln \rho
   + 2 (1-\rho) \ln (1-\rho) - 2\rho \ln 4],
\]
%\end{multline*}
which gives $\rho^* = 0.634641$ and $f_2(\rho^*) = 0.661355$.
The approximate formula of  Lin and Lai is 
\[
  f_2^{\text{LL}}(\rho) = 
  -\frac{\rho}{2} \ln \frac{\rho}{2} - 0.9030 (1-\rho) \ln(1-\rho)
  -0.05645 \rho
\]
which gives $\rho^* = 0.638057$ and $f_2(\rho^*) = 0.66282235$.
Although the two formulas are quite simple, they give 
effective approximation with respect to  $\rho^*$ and $f_2(\rho^*)$.

\begin{comment}

\begingroup
\squeezetable
\begin{table}
\caption{\label{T:}}
\begin{ruledtabular}
\begin{tabular}{lll}
19/30      & 0.63333	    &   0.662761073 \\
45/71      & 0.63380	    &   0.6627681 \\
26/41      & 0.63415	    &   0.662772835 \\
59/93      & 0.63441	    &   0.66277617 \\
33/52      & 0.63462	    &   0.662778631 \\
40/63      & 0.63492	    &   0.662782015 \\
47/74      & 0.63514	    &   0.66278421 \\
61/96      & 0.63542	    &   0.662786858 \\
103/162    & 0.63580	    &   0.66279006 \\
7/11       & 0.63636	    &   0.662793850 \\
107/168    & 0.63690	    &   0.662796516 \\
58/91      & 0.63736	    &   0.66279802 \\
51/80      & 0.63750	    &   0.66279833 \\
44/69      & 0.63768	    &   0.662798649 \\
37/58      & 0.63793	    &   0.662798912 \\
67/105     & 0.63810	    &   0.66279897 \\
30/47      & 0.63830	    &   0.662798922 \\
53/83      & 0.63855	    &   0.6627987 \\
23/36      & 0.63889	    &   0.662798001 \\
62/97      & 0.63918	    &   0.6627971 \\
39/61      & 0.63934	    &   0.662796501 \\
55/86      & 0.63953	    &   0.66279567 \\
16/25      & 0.64000	    &   0.662793131 \\
57/89      & 0.64045	    &   0.66279 \\
41/64      & 0.64062	    &   0.66278859 \\
41/64      & 0.64062	    &   0.66278859 \\
25/39      & 0.64103	    &   0.662785 \\
34/53      & 0.64151	    &   0.662779939 \\
43/67      & 0.64179	    &   0.6627766 \\
70/109     & 0.64220	    &   0.66277135 \\
9/14       & 0.64286	    &   0.662761744 \\
\end{tabular}
\end{ruledtabular}
\end{table}
\endgroup

\end{comment}

\section{Comparison with Baxter's results} \label{S:Baxter}
Using variational approach, Baxter calculated $h_2(x)$ 
(which is $\ln \kappa$ using his notation)
and $\theta(x)$ (which is $2 \rho$ by his notation)
for different values of dimer activity $x$ ($s^2$ by his notation).
By using Eqs. \ref{E:theta} and \ref{E:h2_theta},
we can compare our results
with Baxter's results in his Table II.
For each of his data point at a dimer activity $x$, 
we calculate $f_2(\rho)$ with $\rho = \theta(x)$.
Then his $h_2(x)$ is converted to 
$f_2^B(\rho) = h_2(x) - \frac{\rho} {2} \ln(x)$.
The comparisons are shown in Table \ref{T:baxter}.
It should be pointed out that in Baxter's data, 
extrapolation is used for the sequence to obtain
$\theta(x)$ and $h_2(x)$
when $x^{-1/2}$ is small 
($x^{-1/2} < 0.3$ for $h_2(x)$ and $x^{-1/2} < 0.5$ for $\theta(x)$), 
while no extrapolation is used
in our data: we only look at the digits that have been converged
for $n=16$ and $n=17$.
Although the extrapolation used in Baxter's data makes
the comparison less direct, we still see that
the agreement is excellent.
It seems that Baxter's method converges faster for $\rho$ very close to $1$
(again the extrapolation factor has to be considered),
and our method is more accurate when $\rho$ is not too close to $1$. 
As in Section \ref{S:cylinder},
we only present the raw data here.
If extrapolation is used,
more correct digits can be obtained.

\begin{table}
\caption{
Comparison with Baxter's results.
Numbers in square brackets indicate the next digits
for $n=16$ (upper bound) and $n=17$ (lower bound).
\label{T:baxter}}
\begin{ruledtabular}
\begin{tabular}{llll}
$x^{-1/2}$  & $\rho$  & $f_2(\rho)$  &  $f_2^B(\rho)$  \\
\hline
0.00 & 1.0            & 0.29[0..3]           & 0.291557 \\
0.02 & 0.994176       & 0.319[2..8]          & 0.3194631 \\
0.05 & 0.9836216      & 0.355[0..2]          & 0.35510683 \\
0.10 & 0.96456376     & 0.4047[5..8]         & 0.404771005 \\
0.20 & 0.924706050    & 0.4810[8..9]         & 0.4810887477 \\
0.30 & 0.8846581140   & 0.536892[1..4]       & 0.5368922350 \\
0.40 & 0.8453815864   & 0.5782845[2..9]      & 0.5782845477 \\
0.50 & 0.8072764728   & 0.608814[3..4]       & 0.6088143934 \\
0.60 & 0.7705280966   & 0.63085609[6..8]     & 0.6308560970 \\
0.80 & 0.7013863228   & 0.655894637[3..5]    & 0.6558946374 \\
1.00 & 0.6381231092   & 0.6627989728[3..4]   & 0.6627989726 \\
1.50 & 0.5042633294   & 0.6349499289380[4..9]
                                             & 0.6349499290 \\
2.00 & 0.4006451804   & 0.5779686472227[1..4]            
                                             & 0.5779686472 \\
2.50 & 0.3211782498   & 0.5140847735884[4..6] 
                                             & 0.5140847737 \\
3.00 & 0.2603068980   & 0.4528361791290[7..9]
                                             & 0.4528361790 \\
3.50 & 0.2134739142   & 0.3978378948658[1..3] 
                                             & 0.3978378949 \\
4.00 & 0.17715243204  & 0.3499573614350[0..2]
                                             & 0.3499573615 \\
4.50 & 0.14869898092  & 0.3088705309099[2..6]
                                             & 0.3088705306\\
5.00 & 0.126162903820 & 0.273811439807[1..2]
                                             & 0.2738114398 \\

\end{tabular}
\end{ruledtabular}
\end{table}

\section{High \dd near close packing} \label{S:hdd}

It is well known that phase transition for the \md model
only occurs at $\rho=1$ \cite{Heilmann1972}.
Since the close-packed dimer system is at the critical point,
it is interesting to investigate the behavior of the model
when $\rho \rightarrow 1$.
Using the similar computational procedure outlined before,
the following results are obtained at high \dd limit:
\begin{equation} \label{E:f_m_n}
 f_{\infty,n}(\rho) \sim
f_{\infty, n}^{\text{lattice}} (1) +  
 \begin{cases}
   -(1-\rho) \ln(1-\rho) & \text{$n$ is odd}\\
   -\frac{1}{2} (1-\rho) \ln(1-\rho) & \text{$n$ is even}\\
   \end{cases}
\end{equation}
where $f_{\infty, n}^{\text{lattice}} (1)$
is the free energy of close-packed lattice with width $n$,
and is given, based on the boundary condition, 
by Eq. \ref{E:a_0_e} (cylinder lattices) 
or Eq. \ref{E:a_0_e_fb} (lattices with
free boundary condition).
Eq. \ref{E:f_m_n} for $n=1$ is verified from the exact result
as shown in Eq. \ref{E:f1}.
The result is also confirmed for other values of $n$ 
by using the Pemantle and Wilson asymptotic methods for multivariate 
generating function, as described in Section \ref{S:lc}. 
For space limitation these confirmations are not presented in this paper.

The dependence of the asymptotic form of $f_{\infty,n}(\rho)$ 
on the parity of the lattice width $n$
as shown in Eq. \ref{E:f_m_n}
reminds us of the results reported previously
for \md model with fixed number of monomers $v$ \cite{Kong2006b},
in which the coefficient of the logarithmic correction term
of the free energy depends on the parity of the lattice width $n$.
These two results are consistent with each other.
If we substitute the relation $v = (1-\rho) mn$ into Eq. \ref{E:f_m_n},
we will get the logarithmic correction term with coefficient $v$
for odd $n$, and $v/2$ for even $n$.
More discussions about this asymptotic form will be found
in Section \ref{S:diss} (Eq. \ref{E:f_asympt}).

\begin{figure}
  \centering
  \includegraphics[angle=270,width=\columnwidth]{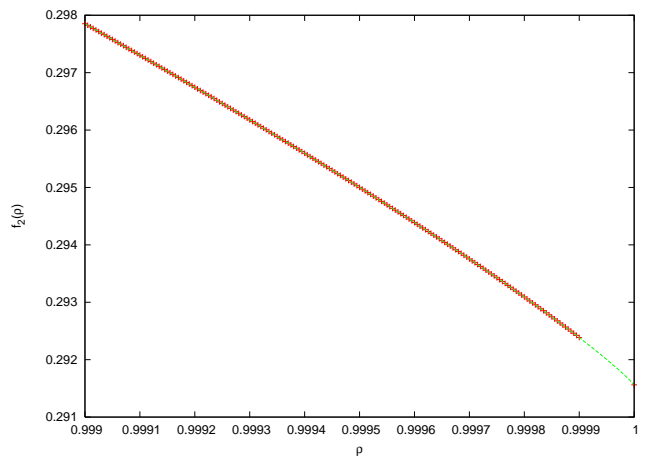}
  \caption{(Color online)
    Fitting result for $f_2(\rho)$ as $\rho \rightarrow 1$. 
    \label{F:f2}}
\end{figure}

We also investigate the behavior of $f_2(\rho)$ (for infinite lattice) 
as $\rho \rightarrow 1$.
Since $f_{\infty,n}(\rho)$ does not converge fast enough
as $\rho \rightarrow 1$ (Table \ref{T:c}), 
we use weighted average of $f_{\infty,16}(\rho)$
and $f_{\infty,17}(\rho)$ as an approximation of $f_2(\rho)$.
The weights are calculated from the exact results at $\rho=1$.
Fitting these data to the following function
\begin{equation} \label{E:f2_fit}
 f_2(\rho) = G/\pi + \frac{\gamma}{2} (1-\rho) \ln{(1-\rho)} + b_1 (1-\rho),
\end{equation}
we obtain $\gamma \approx 1.69775$ and $b_1 \approx 0.427832$.
No other reasonable form of functions other than Eq. (\ref{E:f2_fit})
gives better fit.
Including a term of $(1-\rho)^2$ in Eq. \ref{E:f2_fit} 
leads to only slight changes in
 the values
of $\gamma$ and $b_1$
The data and the fitting result are shown in Figure \ref{F:f2}.

Using the equivalence between  statistical ensembles
discussed in Appendix \ref{S:ee}, we can relate our results
with Gaunt's series expansions \cite{Gaunt1969}.
Plugging $f_2(\rho)$ as in Eq. \ref{E:f2_fit} into
$f(\rho) + \frac{\rho}{2} \ln(x)$ (see Eq. \ref{E:h2}),
differentiating with respect to $\rho$, and solving for $\rho$,
we obtain the average \dd $\theta(x)$ at the activity $x$.
Expressing $x$ as a function of $\theta$, we have
\[
 x = \frac{e^{2 b_1 - \gamma}}{(1-\theta)^\gamma}.
\]
This is in the same form of Eq. (3.7) of Gaunt \cite{Gaunt1969}.
If we put in the values of $\gamma$ and $b_1$,
we can estimate the amplitude $A = \exp(2 b_1 - \gamma) = 0.4308$.
Gaunt obtains through series expansions $\gamma = 1.73 \pm 4$
and $A = 0.3030 \pm 4$, and conjectures that $\gamma = 7/4$.
Our results support the same functional form, and
the numerical values are close to these obtained by Gaunt's
series analysis. 
As for the conjectured value of $\gamma$,
the current data seem to indicate a value lower than $7/4$.
In fact, the data presented here as well as theoretical arguments
(not shown here) indicate that $\gamma = 5/3$.
More discussion on this constant can be found in the next Section.

\section{Discussion}
\label{S:diss}

In Section \ref{S:lc} we show by computational methods that
there is a logarithmic correction term in the free energy
with a coefficient of $-1/2$.
By introducing the newly developed Pemantle and Wilson asymptotic method,
we give a theoretical explanation of this term.
We also demonstrate that this term is not unique to the \md model.
Many statistical lattice models can be casted in the form of
bivariate (or multivariate) generating functions, and
when the two variables are proportional to each other so that
the system is at a fixed ``density'', we will expect
such a universal logarithmic correction term with coefficient of $-1/2$.
We anticipate more applications of this asymptotic method in 
statistical physics in the future. 

The Pemantle and Wilson asymptotic method not only gives
a nice explanation of the logarithmic correction term and its
coefficient found by computational means, 
but also gives \emph{exact} numerical values of 
$f_{\infty, n} (\rho)$ 
for small $n$ (the width
of the lattice strips).
These exact values can be used to check the accuracy of the 
computational method, as already mentioned in Section \ref{S:cylinder}.
In Section \ref{S:lc} we discuss how this can be done.
The denominator $H(x, y)$ of the bivariate generating functions
is derived from the characteristic function of the matrix $M_n$, and the
size of $M_n$ is given by $u_c(n)$ in Section \ref{S:method} for
cylinder lattices, and in Ref. \onlinecite{Kong2006} for lattices
with free boundaries. 
For small $n$, the size of $M_n$ is small enough so that
the characteristic function can be calculated symbolically.
As $n$ increases, however, the size of $M_n$ increases exponentially:
$u_c(17) = 4112$ for cylinder lattice when $n=17$,
and $u_{\text{fb}}(16) = 32896$ when $n=16$.
It is currently impractical to calculate the characteristic functions
symbolically from matrices of such sizes to get $H(x, y)$
of the corresponding bivariate generating functions, 
so the Pemantle and Wilson method cannot be applied
when the width of the lattice $n$ becomes larger.
Even when $H(x, y)$ is available,
it is of the order of thousands or higher,
which will lead to instabilities in the numerical calculations.
The computational method utilized here, however,
can still give important and accurate data
in these situations.

Previously we demonstrated that when the monomer number $v$ or
the dimer number $s$ are fixed, there is also a 
logarithmic correction term in the free energy \cite{Kong2006,Kong2006b}.
When the number of dimers is fixed (low \dd limit), the coefficient
of this term equal to the number of dimers. 
When the number of monomers is fixed (high \dd limit), 
the coefficient, however, depends on the parity of the lattice
width $n$: it equals to $v$ when $n$ is odd, and
$v/2$ when $n$ is even.
In this high \dd limit, as $m \rightarrow \infty$,
\dd $\rho \rightarrow 1$.
In this paper we focus on the situation where the \dd is fixed, 
and find that again there is a logarithmic correction term, 
but this time 
its coefficient equals to $-1/2$ and
does not depend on the parity of the lattice width.
Why does the dependence of the coefficient on the parity
of the lattice width disappear
as \dd $\rho \rightarrow 1$ and the lattice becomes almost completely covered
by the dimers?

This seemingly paradoxical phenomenon can be explained as follows.
When the number of monomer $v$ is fixed and as $m \rightarrow \infty$,
if we can put $v = (1-\rho)mn$ into Eq. \ref{E:f_m_n}, 
then the term of $(1-\rho) \ln(1-\rho)$ leads to a term of $v \ln m /(mn)$
when $n$ is odd, and a term of $v \ln m /(2mn)$
when $n$ is even.  
At the same time, the logarithmic correction term 
with $-1/2$ as coefficient ($-\ln m /(2mn)$,  
the second term in Eq. \ref{E:asympt_md}),
gets canceled out by the a term of $- \ln (1-\rho)/(2mn)$
from the third term in Eq. \ref{E:asympt_md} as $m \rightarrow \infty$ and
$\rho \rightarrow 1$.
Putting Eqs. \ref{E:asympt_md} and \ref{E:f_m_n} together,
we have for finite $n$, as $m \rightarrow \infty$  and
$\rho \rightarrow 1$,
\begin{widetext}
\begin{equation} \label{E:f_asympt}
  f_{m, n}(\rho) \sim
  f_{\infty, n}^{\text{lattice}} (1) +  
  \left\{ 
  \begin{tabular}{l} 
    $-(1-\rho) \ln(1-\rho)$ \\
    $-\frac{1}{2} (1-\rho) \ln(1-\rho) $
  \end{tabular} 
  - \frac{1}{2mn} \ln m 
  - \frac{1}{2mn} \ln (1-\rho)
  \quad
  \begin{tabular}{l} 
    $n$ is odd\\
    $n$ is even\\
  \end{tabular} 
  \right.
\end{equation}
\end{widetext}
This expression can be checked with the explicit formulas
for one-dimensional lattice, Eqs. \ref{E:n1_1} and \ref{E:f1}.
By using the relation $v=(1-\p)mn$, we see from Eq. \ref{E:f_asympt} 
that  as $m \rightarrow \infty$,
when the monomer number $v$ is fixed, 
the dependence of the logarithmic correction term 
on the parity of $n$ comes from the second term in the equation;
the third and fourth terms cancel each other out.
On the other hand 
when the \dd $\rho$ is fixed, the only logarithmic correction term
comes from the third term of Eq. \ref{E:f_asympt}, with coefficient
$-1/2$.

As $n \rightarrow \infty$, $f_2(\rho)$ also has a term of 
$(1-\rho) \ln(1-\rho)$ (Eq. \ref{E:f2_fit}),
whose  coefficient
is estimated as $-0.85$ (Section \ref{S:hdd}).
This value is closer to $-5/6=-0.83$ 
than to the conjectured value $-7/8=-0.875$ by Gaunt 
\cite{Gaunt1969}.
Runnels' result, however, seems to be closer to Gaunt's result 
\cite{Runnels1970}.
It should be recognized that, 
as pointed out previously 
\cite{Nagle1966,Baxter1968,Gaunt1969,Runnels1970,Samuel1980c} as well as
in the present work, the convergence is poorest
when the \dd is near close packing.
%The discrepancy in the estimated values of $\gamma$ may just
%be due to the errors introduced by the slow convergence.
%To get a definite conclusion about 
%the true value of this coefficient, 
%further theoretical and computational 
%developments are needed.
%investigations need to be 
%carried out.
%It should be borne in mind, however, that the convergence 
%is slow when the \dd is near close packing,
%and apprixmate weight factors have to be used in these calculations.
On the other hand,
theoretical calculations underway (not shown here due to space limitation)
indeed 
indicate that this coefficient of $(1-\rho) \ln(1-\rho)$
for the infinite lattice is $-5/6$,
or equivalently, $\gamma = 5/3$.

It is well known
that there is a one to one correspondence
between the Ising model in a rectangular lattice
with zero magnetic field 
and a fully packed dimer model
in a decorated lattice \cite{Kasteleyn1963,Fisher1966}.
By using a similar method \cite{Heilmann1972},
it has been shown that
the Ising model in a non-zero magnetic field,
a well-known unsolved problem in statistical mechanics,
can be mapped to a \md model with \dd $\rho<1$.
The investigation of the \md model near close packing 
is of interest within this context.

As mentioned in Section \ref{S:max},
several authors have applied field theoretic methods
to analyze the \md model (for example, Ref. \onlinecite{Samuel1980c}).
In such studies, the \md problem is expressed as a fermionic field theory.
For close-packed dimer model, the expression is a free field theory
with quadratic action, which is exactly solvable as expected.
For general \md model, the expression is an interacting field theory with a quartic interaction term,  
and self-consistent Hartree approximation is used to improve the
Feynman rules to derive the series expansions.
The transforms that are obtained using these methods 
(such as Eq. \ref{E:sam}, which is similar to Eq. \ref{E:nagle})
make the series expansions converge in the full range of the 
dimer activity.
The accuracy of these calculations, however,
is not comparable to the accuracy of the computational method
reported here, 
%when $\rho$ is not very close to zero, 
possibly due to the limited length of the series expansion.

\appendix

\section{Equivalence of statistical ensembles} \label{S:ee}

Throughout the paper our focus is on the functions $f_{m, n}(\rho)$,
$f_{\infty, n}(\rho)$, or $f_2(\rho)$
at a given \dd $\rho$.
These functions are in essence properties of the canonical ensemble.
In this Appendix we make the connection between $f_2(\rho)$
and the functions of $\theta(x)$ and $h_2(x)$
as defined in Eqs. \ref{E:theta-general} and \ref{E:h2-general},
which are properties of the grand canonical ensemble.
The results are used in 
Section \ref{S:hdd} to compare
the results of near close packing \dd with Gaunt's series analysis, and in
Section \ref{S:Baxter}
to compare our results with those of Baxter, whose calculations are
carried out in terms of $\theta(x)$ and $h_2(x)$ \cite{Baxter1968}.

Suppose at $\rho = \rho^*$, the summand $a_{m,n}(\rho) x^{mn\rho/2}$ 
in Eq.~\ref{E:Z_rho} reaches its maximum.
By using the standard thermodynamic equivalence between 
different statistical ensembles
\cite[Chap. 4]{Hill1987},
we have
\begin{align} \label{E:h2}
 h_2(x) &= \lim_{m,n \rightarrow \infty} \frac{\ln Z_{m,n}(x)}{mn} \notag \\
 &= \lim_{m,n \rightarrow \infty} 
 \frac{\ln \sum_{0 \le \rho \le 1} a_{m,n}(\rho) x^{mn\rho/2}}{mn} \notag \\
 &= \lim_{m,n \rightarrow \infty} 
 \frac{\ln  a_{m,n}(\rho^*) x^{mn\rho^*/2}}{mn} \notag \\
 &= f_2(\rho^*) + \frac{\rho^*}{2} \ln x.
\end{align}
In other words, if we define $F_2(\rho, x) = f_2(\rho) + \frac{\rho}{2} \ln x$,
then
\begin{equation} \label{E:h2_ee}
 h_2(x) = \max_{0 \le \rho \le 1} ( f_2(\rho) + \frac{\rho}{2} \ln x)
 = \max_{0 \le \rho \le 1} F_2(\rho, x).
\end{equation}
As a special case, the \md constant is the maximum of the function 
$f_2(\rho)$ by setting $x=1$
\begin{equation} \label{E:h2_1_ee}
 h_2 = \max_{0 \le \rho \le 1}  f_2(\rho)  = f_2(\rho^*).
\end{equation}  
The connection for the average dimer coverage can also be obtained 
by using Eqs. \ref{E:theta-general} and \ref{E:h2} as
\begin{equation} \label{E:theta}
 \theta(x) = \lim_{m,n \rightarrow \infty} \theta_{m,n}(x)
 = \lim_{m,n \rightarrow \infty} 
 \frac{2}{mn} \frac{ \pd \ln Z_{m,n}(x) } {\pd \ln x} 
 = \rho^*(x)
\end{equation}  
with the understanding that at $\rho^*$,
$F_2(\rho, x) = f_2(\rho) + \frac{\rho}{2} \ln x$,
not $f_2(\rho)$, 
reaches its maximum.
Substituting Eq. \ref{E:theta} into Eq. \ref{E:h2}, we obtain
\begin{equation} \label{E:h2_theta}
  h_2(x) = f_2( \theta(x) ) + \frac{\theta(x)}{2} \ln x .
\end{equation}  

In Section \ref{S:max}, 
the excellent agreement of our result of $f_2(\rho^*)$ with the result on 
$h_2$ of Friedland and Peled \cite{FriedlandP05}
has already been demonstrated.
In Ref. \onlinecite{FriedlandP05}, what is calculated is actually
$h_2$. Eq. \ref{E:h2_1_ee} makes it possible to compare our result
with that of Friedland and Peled.
Eq. \ref{E:h2_1_ee} is proved as a theorem for the specific \md problem
in Ref. \onlinecite{FriedlandP05} as Theorem 4.1.

Since there is an analytical solution to the  one-dimensional lattice
problem,
Eqs. \ref{E:h2} and \ref{E:theta} can be confirmed
for the one-dimensional lattice
by explicit calculations, as shown in Appendix \ref{S:1d}.

\section{Explicit results for one-dimensional lattice} \label{S:1d}

In this Appendix we summarize some exact results
for the one-dimensional lattice ($n=1$) which are useful to compare and
check the results for lattices with width $n > 1$.
When $n=1$, the problem is a special case of the so-called ``parking problem''
in one-dimensional lattice
in which a $k$-mer covers $k$ consecutive lattice sites in a non-overlapping
way. Various methods exist which lead to closed form solutions to the general
case of interacting $k$-mers (for example, see Refs.  
\ifthenelse{\equal{\combine}{false}}{
\onlinecite{Zasedatelev1971,DiCera1996,Kong2001}}
{\onlinecite{Zasedatelev1971,DiCera1996}}).
For the \md model, $k=2$ and there is no interaction between
the dimers.
The number of ways to
put $s$ dimers in the $m \times 1$ lattice is known as
\begin{equation} \label{E:d1}
 a_{m, 1}(s) = \binom{m-s}{s} .
\end{equation}
From this expression, in the next subsection we derive the 
asymptotic expression of the
free energy by using the traditional method.
As an illustration, later we also give an explicit demonstration of 
Pemantle and Wilson's asymptotic method as it is applied
to the bivariate generating function.

\subsection{Canonical ensemble}

From the explicit expression of Eq. \ref{E:d1},
 we can get the asymptotic expression of the
free energy by using Stirling formula when $0 < \rho < 1$:
\begin{widetext}
\begin{align}
  & f_{m,1}(\rho) = \frac{\ln  a_{m, 1}(s)}{m} 
  = f_{\infty,1}(\rho) 
  - \frac{1}{2m} \ln m 
  + \frac{1}{2m} 
  %\left[ 
  \ln \frac{2-\rho}{\rho (1-\rho)} 
  %\right] 
  \notag \\
   &
  + \sum_{j=1}^{\infty} \frac{2^{2j-2} B_{2j}}{ j(2j-1) m^{2j}}
  \left[  \frac{1}{(2-\rho)^{2j-1}} - \frac{1}{\rho^{2j-1}} 
    -\frac{1}{2^{2j-1} (1-\rho)^{2j-1}}
      \right] \label{E:n1_1} 
  \\
   &
   = f_{\infty,1}(\rho) 
   - \frac{1}{2m} \ln (m+1) \label{E:n1_2} 
   + \frac{1}{2m} 
   %\left[ 
   \ln \frac{2-\rho}{\rho (1-\rho)} 
   %\right] 
   \notag \\&
   + \sum_{j=1}^{\infty} \frac{1}{m^{2j}} 
   \left[ \frac{2^{2j-2} B_{2j}}{ j(2j-1) } \left(
     \frac{1}{(2-\rho)^{2j-1}} - \frac{1}{\rho^{2j-1}} 
     -\frac{1}{2^{2j-1} (1-\rho)^{2j-1}}
     \right) + \frac{1}{2(2j-1)} \right] 
   %\notag \\&
   -\sum_{j=1}^{\infty} \frac{1}{4j m^{2j+1}} 
\end{align}
\end{widetext}
where
\begin{equation} \label{E:r1}
 f_{\infty,1}(\rho) = (1 - \frac{\rho}{2}) \ln (1 - \frac{\rho}{2})
 -\frac{\rho}{2} \ln \frac{\rho}{2}
 - (1-\rho) \ln(1-\rho)
\end{equation}
and $B_{2j}$ are the Bernoulli numbers.

From Eqs. \ref{E:n1_1} or \ref{E:n1_2}
it is evident that for $n=1$, 
the coefficient of the logarithmic correction
term is $\ell = -1/2$ for $0 < \rho < 1$.

The asymptotic expression of $f_{\infty,1}(\rho)$ (Eq. \ref{E:r1}) 
at $\rho=1$ is given by
\begin{equation} \label{E:f1}
 f_{\infty,1}(\rho) \sim
 -(1-\rho) \ln (1-\rho) - (\ln2-1)(1-\rho) 
 - \sum_{i=1} \frac{(1-\rho)^{2i+1}}{2i(2i+1)}
\end{equation}
From this asymptotic expression we can see that the coefficient of
$(1-\rho) \ln (1-\rho)$ is exactly $-1$, 
as in Eq. \ref{E:f_m_n} for odd values of $n$.
By combining Eqs. \ref{E:n1_1} and \ref{E:f1} together
we confirm Eq. \ref{E:f_asympt} for $n=1$
at high \dd limit.

\subsection{Grand canonical ensemble}

In this section we calculate various quantities associated
with the grand canonical ensemble.
The configurational grand canonical partition function (Eq. \ref{E:gpf}) is
\[
 Z_{m,1}(x) = \sum_{s=0}^{m/2} \binom{m-s}{s} x^s
\]
To get a closed form of $Z_{m,1}(x)$, we use the WZ method
(Wilf-Zeilberger) \cite{Petkovsek1996} 
to get the following recurrence of $Z_{m,1}(x)$
\[
 x Z_{m,1}(x) + Z_{m+1,1}(x) - Z_{m+2,1}(x) = 0,
\]
from which we obtain the closed form solution as
\begin{equation} \label{E:Z_1d}
 Z_{m,1}(x) = 
 \frac{1}{\sqrt{1+4x}}
 \left[ \beta_1^{m+1} - \beta_2^{m+1}
   \right]
\end{equation}
where
\[
 \beta_{1,2} = \frac{1 \pm \sqrt{1+4x}}{2}.
\]
To calculate $\theta(x)$ using Eq. \ref{E:theta-general},
we need to evaluate the sum 
\[
S(m) = \sum_{s=0}^{m/2} \binom{m-s}{s} s x^s.
\]
Again by using WZ method, we obtain the following recurrence for
$S(m)$
\[
 (m+2) x S(m) + (m+1) S(m+1) - m S(m+2) = 0.
\]
To solve this recurrence, we use the generating function of $S(m)$:
$G_S(z) = \sum_{m} S(m) z^m$, and get $G_S(z)$ from the recurrence as
\[
 G_S(z) = \frac{x z^2}{ (1-z-xz^2)^2}.
\]
From $G_S(z)$ a closed form expression of $S(m)$ can be found as
%\begin{widetext}
\begin{equation} \label{E:F_1d}
 S(m) = \frac{x}{1+4x}
 \left[
   (m - \frac{1}{\sqrt{1+4x}}) \beta_1^{m}
   + (m + \frac{1}{\sqrt{1+4x}}) \beta_2^{m}
   \right]
\end{equation}
%\end{widetext}
Substituting Eqs. \ref{E:Z_1d} and \ref{E:F_1d} into Eq. \ref{E:theta-general},
we obtain
\begin{equation} \label{E:theta_1d}
 \theta_1(x) = 1 - \frac{1}{\sqrt{1+4x}}
\end{equation}
Using Eq. \ref{E:Z_1d} we can calculate $h_1(x)$ as
\begin{equation} \label{E:h_1}
 h_1(x) = \lim_{m \rightarrow \infty} \frac{\ln Z_{m,1}(x)}{m}
 = \ln \frac{1+\sqrt{1+4x}}{2}
\end{equation}

It is known that there are multiple methods to solve
the one-dimensional lattice model. For example, transform matrix method 
can also be used \cite{Wu2006c}. In this case, the transform matrix is
\[
 T_1 = 
 \begin{bmatrix} 
   0 & x \\
   1 & 1
 \end{bmatrix},
\]
whose eigenvalues are $\lambda_{1,2} = (1 \pm \sqrt{1+4x})/2$.
As $m \rightarrow \infty$, $Z_{m,1}(x) \sim \lambda_{1}^m$, so we obtain
$h_1(x)$ as Eq. \ref{E:h_1}.
From $\theta_1(x) = 2 \pd \ln \lambda_{1} / \pd \ln x$ we obtain 
Eq. \ref{E:theta_1d}.

\subsection{Equivalence of statistical ensembles}

The confirmation of the equivalence of ensembles
for the special case of $x=1$ (Eq. \ref{E:h2_1_ee})
has been done in Ref. \onlinecite{FriedlandP05}.
Here we carry out the explicit calculations for the
general case of arbitrary dimer activity $x$.

If we take the derivative of function
$F_1(\rho, x) = f_1(\rho) + \frac{\rho}{2} \ln x$,
where $f_1(\rho) = f_{\infty, 1}(\rho)$ is given in Eq. \ref{E:r1},
solve for $\rho$, and retain only the solution in $[0, 1]$, we have
\[
 \rho^* = 1  - \frac{1}{\sqrt{1+4x}}
\]
which is the same as Eq. \ref{E:theta_1d}.
If we put the value of $\rho^*$ into $F_1(\rho, x)$,
we obtain the maximum of $F_1(\rho^*, x)$:
\[
\max_{0 \le \rho \le 1} ( f_1(\rho) + \frac{\rho}{2} \ln x)
 = \ln \frac{1+\sqrt{1+4x}}{2} 
 = h_1(x)
\]

\subsection{Application of Pemantle and Wilson asymptotic method
to the bivariate generating function}

The bivariate generating function of Eq, \ref{E:bgf} 
can also be obtained in multiple ways,
for example by direct summation of Eq. \ref{E:Z_1d},
or by using the characteristic function of $M_1$,
or by using Eq. (23) of Ref. \onlinecite{DiCera1996} 
(by setting the interaction parameter $\sigma = 1$ and the size of $k$-mer as 2),
to get
\begin{equation} \label{E:F_1}
 F_1(x, y) =  \frac{1}{1-y-xy^2}.
\end{equation}
Here $G(x, y) = 1$ and $H(x, y) = 1-y-xy^2$.
Solving the two equations in Eq. \ref{E:asympt_dir} we get
$(x_0, y_0)$ as 
$(x_0 = s(m-s)/(m-2s)^2, y_0 = (m-2s)/(m-s) )$.
Substituting $(x_0, y_0)$ into Eq. \ref{E:asympt} we obtain
\[
 a_{m,1} \sim \frac{1}{\sqrt{ 2 \pi}} 
(m-s)^{m-s + 1/2} 
(m-2s)^{-m+2s-1/2} (s)^{-s-1/2}.
\]
By putting $s = \rho m /2$, the first three terms of Eq. \ref{E:n1_1}
are recovered, including the term of logarithmic correction 
$-\ln m/(2m)$. Higher order terms can also be obtained if more terms of
the asymptotic expressions are used \cite{Pemantle2002}.

\ifthenelse{\equal{\combine}{true}}{

}
{
\bibliography{md,md_1}
}

\end{document}

%% file: latex_table_v2.1.tex
	& 1/4 &	 1/2 &	 3/4 &	 1 &	 1  	\\
\hline
1 &	0.358851778502358&	0.477385626221110&	0.420632291880785&	0.000000000000000&		\\

1 &	0.358851778501632 (113) &	0.477385626220963 (226) &	0.420632291880650 (113) &	3.6259082842339e-31 (451) &	0.000000000000000 \\
2 &	0.443539035661245 (226) &	0.643863506776599 (451) &	0.675072579831534 (226) &	0.440686793509790 (901) &	0.440686793509772 \\
3 &	0.441243226869578 (113) &	0.632058256526847 (226) &	0.634554086596250 (113) &	0.261133206162104 (451) &	0.261133206162069 \\
4 &	0.441350608415009 (451) &	0.633331866235995 (901) &	0.641840174628945 (451) &	0.329239474231224 (901) &	0.329239474231204 \\
5 &	0.441345086182334 (113) &	0.633177665529326 (226) &	0.640045538037963 (113) &	0.280932225367582 (451) &	0.280932225367553 \\
6 &	0.441345392065621 (226) &	0.633198099780748 (451) &	0.640485680552428 (226) &	0.307299539523143 (901) &	0.307299539523125 \\
7 &	0.441345374298049 (113) &	0.633195220523869 (226) &	0.640363389854116 (113) &	0.286180041989361 (451) &	0.286180041989328 \\
8 &	0.441345375366775 (901) &	0.633195644943681 (901) &	0.640398267527096 (901) &	0.300105275372022 (901) &	0.300105275372003 \\
9 &	0.441345375300735 (113) &	0.633195580174568 (226) &	0.640387826199450 (113) &	0.288315256713912 (451) &	0.288315256713877 \\
10 &	0.441345375304906 (226) &	0.633195590329820 (451) &	0.640391026472015 (226) &	0.296935925720006 (901) &	0.296935925719986 \\
11 &	0.441345375304640 (113) &	0.633195588702860 (226) &	0.640390021971380 (113) &	0.289391267149380 (451) &	0.289391267149350 \\
12 &	0.441345375304658 (451) &	0.633195588968099 (901) &	0.640390342494518 (451) &	0.295260881552885 (901) &	0.295260881552868 \\
13 &	0.441345375304658 (113) &	0.633195588924235 (226) &	0.640390238745032 (113) &	0.290008735546277 (451) &	0.290008735546247 \\
14 &	0.441345375304656 (196) &	0.633195588931575 (391) &	0.640390272712621 (196) &	0.294265803657058 (781) &	0.294265803657028 \\
15 &	0.441345375304652 (71) &	0.633195588930329 (143) &	0.640390261482688 (71) &	0.290395631458758 (285) &	0.290395631458698 \\
16 &	0.441345375304640 (375) &	0.633195588930530 (375) &	0.640390265226077 (375) &	0.293625491565320 (375) &	0.293625491565145 \\
17 &	0.441345375304620 (28) &	0.633195588930470 (57) &	0.640390263969286 (28) &	0.290653983951606 (113) &	0.290653983951281 \\

%% file: latex_table_fb_v2.1.tex
	& 1/4 &	 1/2 &	 3/4 &	 1 &	 1 	\\
\hline
2 &	0.406768721898144 (101) &	0.567460205873414 (201) &	0.550618824275690 (101) &	0.240605912529824 (401) &	0.240605912529802 \\
3 &	0.418805029581931 (50) &	0.589202338098224 (101) &	0.577814537070212 (50) &	0.219492982820793 (201) &	0.219492982820803 \\
4 &	0.424677898377694 (201) &	0.600481083876114 (401) &	0.593860234282314 (201) &	0.260998208772619 (401) &	0.260998208772539 \\
5 &	0.428121453697918 (50) &	0.607125402184205 (101) &	0.603150396985283 (50) &	0.252922288709197 (201) &	0.252922288709162 \\
6 &	0.430386238446347 (101) &	0.611530695170404 (201) &	0.609386552832152 (101) &	0.269862305348313 (401) &	0.269862305348238 \\
7 &	0.431988836086132 (50) &	0.614662382427737 (101) &	0.613828224552787 (50) &	0.265557149993036 (201) &	0.265557149992917 \\
8 &	0.433182588323077 (401) &	0.617003233867274 (401) &	0.617157805951205 (401) &	0.274751610011806 (401) &	0.274751610011700 \\
9 &	0.434106231271596 (50) &	0.618819152717284 (101) &	0.619745522952440 (50) &	0.272072662436541 (201) &	0.272072662436436 \\
10 &	0.434842114982272 (101) &	0.620268892851619 (201) &	0.621814606701445 (101) &	0.277844105572086 (401) &	0.277844105571997 \\
11 &	0.435442204742419 (50) &	0.621453058127923 (101) &	0.623506740930563 (50) &	0.276016066623932 (201) &	0.276016066623911 \\
12 &	0.435940910510948 (201) &	0.622438494443121 (401) &	0.624916337018327 (201) &	0.279975752031904 (401) &	0.279975752031819 \\
13 &	0.436361922501114 (39) &	0.623271352033866 (78) &	0.626108703477498 (39) &	0.278648778924217 (155) &	0.278648778924210 \\
14 &	0.436722083762241 (26) &	0.623984518924941 (51) &	0.627130461956123 (26) &	0.281534000787684 (101) &	0.281534000780413 \\
15 &	0.437033697160078 (13) &	0.624602065315795 (26) &	0.628015783739589 (13) &	0.280526932170974 (51) &	0.280526932164772 \\
16 &	0.437305958542365 (44) &	0.625142013068189 (44) &	0.628790285827699 (44) &	0.282722754819597 (44) &	0.282722752409010 \\

%% file: latex_table_v6.1.tex
	&0.63812309 &	0.63812310 &	0.63812311 &	0.63812312 	\\
\hline
1 &	0.470643631091106 (880) &	0.470643628559868 (880) &	0.470643626028629 (880) &	0.470643623497390 (880) \\
2 &	0.683451694063943 (901) &	0.683451695019491 (901) &	0.683451695975038 (901) &	0.683451696930585 (901) \\
3 &	0.659839104062378 (901) &	0.659839103873019 (901) &	0.659839103683659 (901) &	0.659839103494298 (901) \\
4 &	0.663319985040839 (901) &	0.663319985089007 (901) &	0.663319985137175 (901) &	0.663319985185343 (901) \\
5 &	0.662701144811933 (901) &	0.662701144800592 (901) &	0.662701144789251 (901) &	0.662701144777910 (901) \\
6 &	0.662818978977777 (901) &	0.662818978980627 (901) &	0.662818978983477 (901) &	0.662818978986327 (901) \\
7 &	0.662794695257766 (901) &	0.662794695257048 (901) &	0.662794695256327 (901) &	0.662794695255611 (901) \\
8 &	0.662799924786436 (901) &	0.662799924786622 (901) &	0.662799924786807 (901) &	0.662799924786992 (901) \\
9 &	0.662798754939549 (901) &	0.662798754939501 (901) &	0.662798754939453 (901) &	0.662798754939404 (901) \\
10 &	0.662799023857733 (901) &	0.662799023857746 (901) &	0.662799023857758 (901) &	0.662799023857771 (901) \\
11 &	0.662798960670265 (901) &	0.662798960670262 (901) &	0.662798960670259 (901) &	0.662798960670256 (901) \\
12 &	0.662798975775941 (901) &	0.662798975775943 (901) &	0.662798975775944 (901) &	0.662798975775944 (901) \\
13 &	0.662798972113454 (901) &	0.662798972113453 (901) &	0.662798972113445 (901) &	0.662798972113451 (901) \\
14 &	0.662798973011855 (781) &	0.662798973011855 (781) &	0.662798973011855 (781) &	0.662798973011855 (781) \\
15 &	0.662798972789303 (570) &	0.662798972789304 (570) &	0.662798972789304 (570) &	0.662798972789303 (570) \\
16 &	0.662798972844882 (375) &	0.662798972844882 (375) &	0.662798972844883 (375) &	0.662798972844883 (375) \\
17 &	0.662798972830869 (226) &	0.662798972830871 (226) &	0.662798972830870 (226) &	0.662798972830869 (226) \\

%% file: rho.bbl
\begin{thebibliography}{99}
\expandafter\ifx\csname natexlab\endcsname\relax\def\natexlab#1{#1}\fi
\expandafter\ifx\csname bibnamefont\endcsname\relax
  \def\bibnamefont#1{#1}\fi
\expandafter\ifx\csname bibfnamefont\endcsname\relax
  \def\bibfnamefont#1{#1}\fi
\expandafter\ifx\csname citenamefont\endcsname\relax
  \def\citenamefont#1{#1}\fi
\expandafter\ifx\csname url\endcsname\relax
  \def\url#1{\texttt{#1}}\fi
\expandafter\ifx\csname urlprefix\endcsname\relax\def\urlprefix{URL }\fi
\providecommand{\bibinfo}[2]{#2}
\providecommand{\eprint}[2][]{\url{#2}}

\bibitem[{\citenamefont{Fowler and Rushbrooke}(1937)}]{Fowler1937}
\bibinfo{author}{\bibfnamefont{R.~H.} \bibnamefont{Fowler}} \bibnamefont{and}
  \bibinfo{author}{\bibfnamefont{G.~S.} \bibnamefont{Rushbrooke}},
  \bibinfo{journal}{Trans. Faraday Soc.} \textbf{\bibinfo{volume}{33}},
  \bibinfo{pages}{1272} (\bibinfo{year}{1937}).

\bibitem[{\citenamefont{Kasteleyn}(1961)}]{Kasteleyn1961}
\bibinfo{author}{\bibfnamefont{P.~W.} \bibnamefont{Kasteleyn}},
  \bibinfo{journal}{Physica} \textbf{\bibinfo{volume}{27}},
  \bibinfo{pages}{1209} (\bibinfo{year}{1961}).

\bibitem[{\citenamefont{Temperley and Fisher}(1961)}]{Temperley1961}
\bibinfo{author}{\bibfnamefont{H.~N.~V.} \bibnamefont{Temperley}}
  \bibnamefont{and} \bibinfo{author}{\bibfnamefont{M.~E.}
  \bibnamefont{Fisher}}, \bibinfo{journal}{Philos. Mag.}
  \textbf{\bibinfo{volume}{6}}, \bibinfo{pages}{1061} (\bibinfo{year}{1961});
%
%\bibitem[{\citenamefont{Fisher}(1961)}]{Fisher1961}
\bibinfo{author}{\bibfnamefont{M.~E.} \bibnamefont{Fisher}},
  \bibinfo{journal}{Phys. Rev.} \textbf{\bibinfo{volume}{124}},
  \bibinfo{pages}{1664} (\bibinfo{year}{1961}).

\bibitem[{\citenamefont{Tzeng and Wu}(2003)}]{Tzeng2003}
\bibinfo{author}{\bibfnamefont{W.-J.} \bibnamefont{Tzeng}} \bibnamefont{and}
  \bibinfo{author}{\bibfnamefont{F.~Y.} \bibnamefont{Wu}}, \bibinfo{journal}{J.
  Stat. Phys.} \textbf{\bibinfo{volume}{110}}, \bibinfo{pages}{671}
  (\bibinfo{year}{2003}).

\bibitem[{\citenamefont{Wu}(2006{\natexlab{a}})}]{Wu2006}
\bibinfo{author}{\bibfnamefont{F.~Y.} \bibnamefont{Wu}},
  \bibinfo{journal}{Phys. Rev. E} \textbf{\bibinfo{volume}{74}},
  \bibinfo{pages}{020104} (\bibinfo{year}{2006}{\natexlab{a}});
%
%\bibitem[{\citenamefont{Wu}(2006{\natexlab{b}})}]{Wu2006b}
%\bibinfo{author}{\bibfnamefont{F.~Y.} \bibnamefont{Wu}},
%  \bibinfo{journal}{Phys. Rev. E} \textbf{\bibinfo{volume}{74}},
%  \bibinfo{pages}{039907} (\bibinfo{year}{2006}{\natexlab{b}}).
\textbf{\bibinfo{volume}{74}}, 
\bibinfo{pages}{039907(E)}
  (\bibinfo{year}{2006}).

\bibitem[{\citenamefont{Garey and Johnson}(1979)}]{Garey1979}
\bibinfo{author}{\bibfnamefont{M.~R.} \bibnamefont{Garey}} \bibnamefont{and}
  \bibinfo{author}{\bibfnamefont{D.~S.} \bibnamefont{Johnson}},
  \emph{\bibinfo{title}{Computers and Intractability, A Guide to the Theory of
  {NP}-Completeness}} (\bibinfo{publisher}{W.H. Freeman and Company},
  \bibinfo{address}{New York}, \bibinfo{year}{1979});
%
%\bibitem[{\citenamefont{Welsh}(1993)}]{Welsh1993}
\bibinfo{author}{\bibfnamefont{D.~J.~A.} \bibnamefont{Welsh}},
  \emph{\bibinfo{title}{Complexity: Knots, Colourings, and Counting}}, vol.
  \bibinfo{volume}{186} of \emph{\bibinfo{series}{London Mathematical Society
  Lecture Note Series}} (\bibinfo{publisher}{Cambridge University Press},
  \bibinfo{year}{1993});
%
%\bibitem[{\citenamefont{Mertens}(2002)}]{Mertens2002}
\bibinfo{author}{\bibfnamefont{S.}~\bibnamefont{Mertens}},
  \bibinfo{journal}{Computing in Science and Engineering}
  \textbf{\bibinfo{volume}{4}}, \bibinfo{pages}{31} (\bibinfo{year}{2002}).

\bibitem[{\citenamefont{Jerrum}(1987)}]{Jerrum1987}
\bibinfo{author}{\bibfnamefont{M.}~\bibnamefont{Jerrum}}, \bibinfo{journal}{J.
  Stat. Phys.} \textbf{\bibinfo{volume}{48}}, \bibinfo{pages}{121}
  (\bibinfo{year}{1987});
%
%\bibitem[{\citenamefont{Jerrum}(1990)}]{Jerrum1990}
%\bibinfo{author}{\bibfnamefont{M.}~\bibnamefont{Jerrum}}, \bibinfo{journal}{J.
%  Stat. Phys.} \textbf{\bibinfo{volume}{59}}, \bibinfo{pages}{1087}
%  (\bibinfo{year}{1990}).
\textbf{\bibinfo{volume}{59}}, 
\bibinfo{pages}{1087(E)}
  (\bibinfo{year}{1990}).

\bibitem[{\citenamefont{Nagle}(1966)}]{Nagle1966}
\bibinfo{author}{\bibfnamefont{J.~F.} \bibnamefont{Nagle}},
  \bibinfo{journal}{Phys. Rev.} \textbf{\bibinfo{volume}{152}},
  \bibinfo{pages}{190} (\bibinfo{year}{1966}).

\bibitem[{\citenamefont{Gaunt}(1969)}]{Gaunt1969}
\bibinfo{author}{\bibfnamefont{D.}~\bibnamefont{Gaunt}},
  \bibinfo{journal}{Phys. Rev.} \textbf{\bibinfo{volume}{179}},
  \bibinfo{pages}{174} (\bibinfo{year}{1969}).

\bibitem[{\citenamefont{Samuel}(1980)}]{Samuel1980c}
\bibinfo{author}{\bibfnamefont{S.}~\bibnamefont{Samuel}}, \bibinfo{journal}{J.
  Math. Phys.} \textbf{\bibinfo{volume}{21}}, \bibinfo{pages}{2820}
  (\bibinfo{year}{1980}).

\bibitem[{\citenamefont{Bondy and Welsh}(1966)}]{Bondy1966}
\bibinfo{author}{\bibfnamefont{J.}~\bibnamefont{Bondy}} \bibnamefont{and}
  \bibinfo{author}{\bibfnamefont{D.}~\bibnamefont{Welsh}},
  \bibinfo{journal}{Proc. Camb. Phil. Soc. Math. Phys. Sci.}
  \textbf{\bibinfo{volume}{62}}, \bibinfo{pages}{503} (\bibinfo{year}{1966});
%
%\bibitem[{\citenamefont{Hammersley}(1968)}]{Hammersley1968}
\bibinfo{author}{\bibfnamefont{J.}~\bibnamefont{Hammersley}},
  \bibinfo{journal}{Proc. Camb. Phil. Soc. Math. Phys. Sci.}
  \textbf{\bibinfo{volume}{64}}, \bibinfo{pages}{455} (\bibinfo{year}{1968});
%
%\bibitem[{\citenamefont{Hammersley and Menon}(1970)}]{Hammersley1970}
\bibinfo{author}{\bibfnamefont{J.}~\bibnamefont{Hammersley}} \bibnamefont{and}
  \bibinfo{author}{\bibfnamefont{V.}~\bibnamefont{Menon}}, \bibinfo{journal}{J.
  Inst. Math. Appl.} \textbf{\bibinfo{volume}{6}}, \bibinfo{pages}{341}
  (\bibinfo{year}{1970});
%
%\bibitem[{\citenamefont{Ciucu}(1998)}]{Ciucu1998}
\bibinfo{author}{\bibfnamefont{M.}~\bibnamefont{Ciucu}}, \bibinfo{journal}{Duke
  Math. J} \textbf{\bibinfo{volume}{94}}, \bibinfo{pages}{1}
  (\bibinfo{year}{1998});
%
%\bibitem[{\citenamefont{Lundow}(2001)}]{Lundow2001}
\bibinfo{author}{\bibfnamefont{P.~H.} \bibnamefont{Lundow}},
  \bibinfo{journal}{Disc. Math.} \textbf{\bibinfo{volume}{231}},
  \bibinfo{pages}{321} (\bibinfo{year}{2001}).

\bibitem[{\citenamefont{Friedland and Peled}(2005)}]{FriedlandP05}
\bibinfo{author}{\bibfnamefont{S.}~\bibnamefont{Friedland}} \bibnamefont{and}
  \bibinfo{author}{\bibfnamefont{U.~N.} \bibnamefont{Peled}},
  \bibinfo{journal}{Adv. Appl. Math.} \textbf{\bibinfo{volume}{34}},
  \bibinfo{pages}{486} (\bibinfo{year}{2005}).

\bibitem[{\citenamefont{Fisher and Stephenson}(1963)}]{Fisher1963}
\bibinfo{author}{\bibfnamefont{M.~E.} \bibnamefont{Fisher}} \bibnamefont{and}
  \bibinfo{author}{\bibfnamefont{J.}~\bibnamefont{Stephenson}},
  \bibinfo{journal}{Phys. Rev.} \textbf{\bibinfo{volume}{132}},
  \bibinfo{pages}{1411} (\bibinfo{year}{1963});
%
%\bibitem[{\citenamefont{Hartwig}(1966)}]{Hartwig1966}
\bibinfo{author}{\bibfnamefont{R.~E.} \bibnamefont{Hartwig}},
  \bibinfo{journal}{J. Math. Phys.} \textbf{\bibinfo{volume}{7}},
  \bibinfo{pages}{286} (\bibinfo{year}{1966}).

\bibitem[{\citenamefont{Heilmann and Lieb}(1972)}]{Heilmann1972}
\bibinfo{author}{\bibfnamefont{O.~J.} \bibnamefont{Heilmann}} \bibnamefont{and}
  \bibinfo{author}{\bibfnamefont{E.~H.} \bibnamefont{Lieb}},
  \bibinfo{journal}{Commun. Math. Phys.} \textbf{\bibinfo{volume}{25}},
  \bibinfo{pages}{190} (\bibinfo{year}{1972}).

\bibitem[{\citenamefont{Gruber and Kunz}(1971)}]{Gruber1971}
\bibinfo{author}{\bibfnamefont{C.}~\bibnamefont{Gruber}} \bibnamefont{and}
  \bibinfo{author}{\bibfnamefont{H.}~\bibnamefont{Kunz}},
  \bibinfo{journal}{Commun. Math. Phys.} \textbf{\bibinfo{volume}{22}},
  \bibinfo{pages}{133} (\bibinfo{year}{1971}).

\bibitem[{\citenamefont{Ferdinand}(1967)}]{Ferdinand1967}
\bibinfo{author}{\bibfnamefont{A.~E.} \bibnamefont{Ferdinand}},
  \bibinfo{journal}{J. Math. Phys.} \textbf{\bibinfo{volume}{8}},
  \bibinfo{pages}{2332} (\bibinfo{year}{1967});
%
%\bibitem[{\citenamefont{Izmailian et~al.}(2003)\citenamefont{Izmailian,
%  Oganesyan, and Hu}}]{IzmailianOH03}
\bibinfo{author}{\bibfnamefont{N.~S.} \bibnamefont{Izmailian}},
  \bibinfo{author}{\bibfnamefont{K.~B.} \bibnamefont{Oganesyan}},
  \bibnamefont{and} \bibinfo{author}{\bibfnamefont{C.~K.} \bibnamefont{Hu}},
  \bibinfo{journal}{Phys. Rev. E} \textbf{\bibinfo{volume}{67}},
  \bibinfo{pages}{066114} (\bibinfo{year}{2003});
%
%\bibitem[{\citenamefont{Izmailian et~al.}(2005)\citenamefont{Izmailian,
%  Priezzhev, Ruelle, and Hu}}]{Izmailian2005}
\bibinfo{author}{\bibfnamefont{N.~S.} \bibnamefont{Izmailian}},
  \bibinfo{author}{\bibfnamefont{V.~B.} \bibnamefont{Priezzhev}},
  \bibinfo{author}{\bibfnamefont{P.}~\bibnamefont{Ruelle}}, \bibnamefont{and}
  \bibinfo{author}{\bibfnamefont{C.-K.} \bibnamefont{Hu}},
  \bibinfo{journal}{Phys. Rev. Lett.} \textbf{\bibinfo{volume}{95}},
  \bibinfo{pages}{260602} (\bibinfo{year}{2005}).

\bibitem[{\citenamefont{Chang}(1939)}]{Chang1939}
\bibinfo{author}{\bibfnamefont{T.}~\bibnamefont{Chang}},
  \bibinfo{journal}{Proc. R. Soc. Series A} \textbf{\bibinfo{volume}{169}},
  \bibinfo{pages}{512} (\bibinfo{year}{1939}).

\bibitem[{\citenamefont{Lin and Lai}(1994)}]{LinL94}
\bibinfo{author}{\bibfnamefont{G.~J.} \bibnamefont{Lin}} \bibnamefont{and}
  \bibinfo{author}{\bibfnamefont{P.~Y.} \bibnamefont{Lai}},
  \bibinfo{journal}{Physica A} \textbf{\bibinfo{volume}{211}},
  \bibinfo{pages}{465} (\bibinfo{year}{1994}).

\bibitem[{\citenamefont{Baxter}(1968)}]{Baxter1968}
\bibinfo{author}{\bibfnamefont{R.~J.} \bibnamefont{Baxter}},
  \bibinfo{journal}{J. Math. Phys.} \textbf{\bibinfo{volume}{9}},
  \bibinfo{pages}{650} (\bibinfo{year}{1968}).

\bibitem[{\citenamefont{Nemirovsky and Coutinho-Filho}(1989)}]{Nemirovsky1989}
\bibinfo{author}{\bibfnamefont{A.~M.} \bibnamefont{Nemirovsky}}
  \bibnamefont{and} \bibinfo{author}{\bibfnamefont{M.~D.}
  \bibnamefont{Coutinho-Filho}}, \bibinfo{journal}{Phys. Rev. A}
  \textbf{\bibinfo{volume}{39}}, \bibinfo{pages}{3120} (\bibinfo{year}{1989});
%
%\bibitem[{\citenamefont{Kenyon et~al.}(1996)\citenamefont{Kenyon, Randall, and
%  Sinclair}}]{KenyonRS96}
\bibinfo{author}{\bibfnamefont{C.}~\bibnamefont{Kenyon}},
  \bibinfo{author}{\bibfnamefont{D.}~\bibnamefont{Randall}}, \bibnamefont{and}
  \bibinfo{author}{\bibfnamefont{A.}~\bibnamefont{Sinclair}},
  \bibinfo{journal}{J. Stat. Phys.} \textbf{\bibinfo{volume}{83}},
  \bibinfo{pages}{637} (\bibinfo{year}{1996});
%
%\bibitem[{\citenamefont{Beichl et~al.}(2001)\citenamefont{Beichl, O'Leary, and
%  Sullivan}}]{BeichlOS01}
\bibinfo{author}{\bibfnamefont{I.}~\bibnamefont{Beichl}},
  \bibinfo{author}{\bibfnamefont{D.~P.} \bibnamefont{O'Leary}},
  \bibnamefont{and} \bibinfo{author}{\bibfnamefont{F.}~\bibnamefont{Sullivan}},
  \bibinfo{journal}{Phys. Rev. E} \textbf{\bibinfo{volume}{64}},
  \bibinfo{pages}{016701} (\bibinfo{year}{2001}).

\bibitem[{\citenamefont{Beichl and Sullivan}(1999)}]{BeichlS99}
\bibinfo{author}{\bibfnamefont{I.}~\bibnamefont{Beichl}} \bibnamefont{and}
  \bibinfo{author}{\bibfnamefont{F.}~\bibnamefont{Sullivan}},
  \bibinfo{journal}{J. Comput. Phys.} \textbf{\bibinfo{volume}{149}},
  \bibinfo{pages}{128} (\bibinfo{year}{1999}).

\bibitem[{\citenamefont{Kasteleyn}(1963)}]{Kasteleyn1963}
\bibinfo{author}{\bibfnamefont{P.~W.} \bibnamefont{Kasteleyn}},
  \bibinfo{journal}{J. Math. Phys.} \textbf{\bibinfo{volume}{4}},
  \bibinfo{pages}{287} (\bibinfo{year}{1963}).

\bibitem[{\citenamefont{Fisher}(1966)}]{Fisher1966}
\bibinfo{author}{\bibfnamefont{M.~E.} \bibnamefont{Fisher}},
  \bibinfo{journal}{J. Math. Phys.} \textbf{\bibinfo{volume}{7}},
  \bibinfo{pages}{1776} (\bibinfo{year}{1966}).

\bibitem[{\citenamefont{Fan and Wu}(1970)}]{Fan1970}
\bibinfo{author}{\bibfnamefont{C.}~\bibnamefont{Fan}} \bibnamefont{and}
  \bibinfo{author}{\bibfnamefont{F.}~\bibnamefont{Wu}}, \bibinfo{journal}{Phys.
  Rev. B} \textbf{\bibinfo{volume}{2}}, \bibinfo{pages}{723}
  (\bibinfo{year}{1970}).

\bibitem[{\citenamefont{McCoy and Wu}(1973)}]{McCoy1973}
\bibinfo{author}{\bibfnamefont{B.~M.} \bibnamefont{McCoy}} \bibnamefont{and}
  \bibinfo{author}{\bibfnamefont{T.~T.} \bibnamefont{Wu}},
  \emph{\bibinfo{title}{The two-dimensional Ising Model}}
  (\bibinfo{publisher}{Harvard University Press}, \bibinfo{address}{Cambridge,
  Massachusetts}, \bibinfo{year}{1973}).

\bibitem[{\citenamefont{Pemantle and Wilson}(2002)}]{Pemantle2002}
\bibinfo{author}{\bibfnamefont{R.}~\bibnamefont{Pemantle}} \bibnamefont{and}
  \bibinfo{author}{\bibfnamefont{M.~C.} \bibnamefont{Wilson}},
  \bibinfo{journal}{J. Comb. Theory Ser. A} \textbf{\bibinfo{volume}{97}},
  \bibinfo{pages}{129} (\bibinfo{year}{2002}).

\bibitem[{\citenamefont{Kong}(1999)}]{Kong1999}
\bibinfo{author}{\bibfnamefont{Y.}~\bibnamefont{Kong}}, \bibinfo{journal}{J.
  Chem. Phys.} \textbf{\bibinfo{volume}{111}}, \bibinfo{pages}{4790}
  (\bibinfo{year}{1999}).

\bibitem[{\citenamefont{Kong}(2006{\natexlab{a}})}]{Kong2006}
\bibinfo{author}{\bibfnamefont{Y.}~\bibnamefont{Kong}}, \bibinfo{journal}{Phys.
  Rev. E} \textbf{\bibinfo{volume}{73}}, \bibinfo{pages}{016106}
  (\bibinfo{year}{2006}{\natexlab{a}}).

\bibitem[{\citenamefont{Kong}(2006{\natexlab{b}})}]{Kong2006b}
\bibinfo{author}{\bibfnamefont{Y.}~\bibnamefont{Kong}}, \bibinfo{journal}{Phys.
  Rev. E} \textbf{\bibinfo{volume}{74}}, \bibinfo{pages}{011102}
  (\bibinfo{year}{2006}{\natexlab{b}}).

\bibitem[{\citenamefont{Granlund}(2006)}]{gmp}
\bibinfo{author}{\bibfnamefont{T.}~\bibnamefont{Granlund}},
  \emph{\bibinfo{title}{GNU MP: The GNU Multiple Precision Arithmetic Library}}
  (\bibinfo{year}{2006}), \bibinfo{note}{version 4.2},
  \urlprefix\url{http://www.swox.com/gmp/}.

\bibitem[{\citenamefont{Calkin and Wilf}(1998)}]{Calkin1998}
\bibinfo{author}{\bibfnamefont{N.~J.} \bibnamefont{Calkin}} \bibnamefont{and}
  \bibinfo{author}{\bibfnamefont{H.~S.} \bibnamefont{Wilf}},
  \bibinfo{journal}{SIAM J. Discrete Math.} \textbf{\bibinfo{volume}{11}},
  \bibinfo{pages}{54} (\bibinfo{year}{1998}).

\bibitem[{\citenamefont{Runnels}(1970)}]{Runnels1970}
\bibinfo{author}{\bibfnamefont{L.}~\bibnamefont{Runnels}}, \bibinfo{journal}{J.
  Math. Phys.} \textbf{\bibinfo{volume}{11}}, \bibinfo{pages}{842}
  (\bibinfo{year}{1970}).

\bibitem[{\citenamefont{Stoer and Bulirsch}(1992)}]{Stoer1980}
\bibinfo{author}{\bibfnamefont{J.}~\bibnamefont{Stoer}} \bibnamefont{and}
  \bibinfo{author}{\bibfnamefont{R.}~\bibnamefont{Bulirsch}},
  \emph{\bibinfo{title}{Introduction to Numerical Analysis}}
  (\bibinfo{publisher}{Springer-Verlag}, \bibinfo{address}{New York, USA},
  \bibinfo{year}{1992}), \bibinfo{edition}{2nd} ed.

\bibitem[{\citenamefont{Press et~al.}(1992)\citenamefont{Press, Teukolsky,
  Vetterling, and Flannery}}]{Press1992}
\bibinfo{author}{\bibfnamefont{W.~H.} \bibnamefont{Press}},
  \bibinfo{author}{\bibfnamefont{S.~A.} \bibnamefont{Teukolsky}},
  \bibinfo{author}{\bibfnamefont{W.~T.} \bibnamefont{Vetterling}},
  \bibnamefont{and} \bibinfo{author}{\bibfnamefont{B.~P.}
  \bibnamefont{Flannery}}, \emph{\bibinfo{title}{Numerical Recipes in C: The
  Art of Scientific Computing}} (\bibinfo{publisher}{Cambridge University
  Press}, \bibinfo{address}{New York, NY, USA}, \bibinfo{year}{1992}).

\bibitem[{\citenamefont{Williams and Kelley}(2004)}]{gnuplot40}
\bibinfo{author}{\bibfnamefont{T.}~\bibnamefont{Williams}} \bibnamefont{and}
  \bibinfo{author}{\bibfnamefont{C.}~\bibnamefont{Kelley}},
  \emph{\bibinfo{title}{Gnuplot: An Interactive Plotting Program}}
  (\bibinfo{year}{2004}), \bibinfo{note}{version 4.0},
  \urlprefix\url{http://www.gnuplot.info}.

\bibitem[{\citenamefont{Marquart}(1963)}]{Marquart1963}
\bibinfo{author}{\bibfnamefont{D.}~\bibnamefont{Marquart}},
  \bibinfo{journal}{J. Soc. Indust. Appl. Math.} \textbf{\bibinfo{volume}{11}},
  \bibinfo{pages}{431} (\bibinfo{year}{1963}).

\bibitem[{\citenamefont{Hammersley}(1966)}]{Hammersley1966}
\bibinfo{author}{\bibfnamefont{J.}~\bibnamefont{Hammersley}}, in
  \emph{\bibinfo{booktitle}{Research Papers in Statistics: Festschrift for J.
  Neyman}}, edited by \bibinfo{editor}{\bibfnamefont{F.}~\bibnamefont{David}}
  (\bibinfo{publisher}{Wiley}, \bibinfo{address}{London},
  \bibinfo{year}{1966}), p. \bibinfo{pages}{125}.

\bibitem[{\citenamefont{Hill}(1987)}]{Hill1987}
\bibinfo{author}{\bibfnamefont{T.~L.} \bibnamefont{Hill}},
  \emph{\bibinfo{title}{Statistical Mechanics : Principles and Applications}}
  (\bibinfo{publisher}{Dover Publications}, \bibinfo{address}{New York, USA},
  \bibinfo{year}{1987}).

\bibitem[{\citenamefont{Zasedatelev et~al.}(1971)\citenamefont{Zasedatelev,
  Gurskii, and Volkenshtein}}]{Zasedatelev1971}
\bibinfo{author}{\bibfnamefont{A.}~\bibnamefont{Zasedatelev}},
  \bibinfo{author}{\bibfnamefont{G.}~\bibnamefont{Gurskii}}, \bibnamefont{and}
  \bibinfo{author}{\bibfnamefont{M.}~\bibnamefont{Volkenshtein}},
  \bibinfo{journal}{Mol. Biol.} \textbf{\bibinfo{volume}{5}},
  \bibinfo{pages}{245} (\bibinfo{year}{1971});
%
%\bibitem[{\citenamefont{Kong}(2001)}]{Kong2001}
\bibinfo{author}{\bibfnamefont{Y.}~\bibnamefont{Kong}}, \bibinfo{journal}{J.
  Phys. Chem. B} \textbf{\bibinfo{volume}{105}}, \bibinfo{pages}{10111}
  (\bibinfo{year}{2001}).

\bibitem[{\citenamefont{Cera and Kong}(1996)}]{DiCera1996}
\bibinfo{author}{\bibfnamefont{E.~D.} \bibnamefont{Cera}} \bibnamefont{and}
  \bibinfo{author}{\bibfnamefont{Y.}~\bibnamefont{Kong}},
  \bibinfo{journal}{Biophys. Chem.} \textbf{\bibinfo{volume}{61}},
  \bibinfo{pages}{107} (\bibinfo{year}{1996}).


\bibitem[{\citenamefont{Petkov\v{s}ek et~al.}(1996)\citenamefont{Petkov\v{s}ek,
  Wilf, and Zeilberger}}]{Petkovsek1996}
\bibinfo{author}{\bibfnamefont{M.}~\bibnamefont{Petkov\v{s}ek}},
  \bibinfo{author}{\bibfnamefont{H.}~\bibnamefont{Wilf}}, \bibnamefont{and}
  \bibinfo{author}{\bibfnamefont{D.}~\bibnamefont{Zeilberger}},
  \emph{\bibinfo{title}{A=B}} (\bibinfo{publisher}{AK Peters, Ltd.},
  \bibinfo{address}{Wellesley, MA, USA}, \bibinfo{year}{1996}).

\bibitem[{\citenamefont{Wu}()}]{Wu2006c}
\bibinfo{author}{\bibfnamefont{F.~Y.} \bibnamefont{Wu}}, \bibinfo{note}{private
  communication}.

\end{thebibliography}
